\documentclass[pdflatex,sn-mathphys-num]{sn-jnl}

\usepackage{graphicx}%
\usepackage{float}
\usepackage{multirow}%
\usepackage{amsmath,amssymb,amsfonts}%
\usepackage{amsthm}%
\usepackage{mathrsfs}%
\usepackage[title]{appendix}%
\usepackage{xcolor}%
\usepackage{textcomp}%
\usepackage{manyfoot}%
\usepackage{booktabs}%
\usepackage{placeins} 
\usepackage{tabularx}
\usepackage{booktabs}
\usepackage{multirow}

\theoremstyle{thmstyleone}%
\theoremstyle{thmstyletwo}%

\theoremstyle{thmstylethree}%

\begin{document}

\title[]{AI-Driven Neural Surrogates for In Silico Design of Cognitive-Affective Neuromodulation Targets}


\author[1]{\fnm{Marco} \sur{Rothermel}}
\author[1,2]{\fnm{Madleen} \sur{Stenger}}
\author[1]{\fnm{Soroush} \sur{Daftarian}}
\author[1]{\fnm{Svenja Jule} \sur{Francke}}
\author[1,2]{\fnm{Bita} \sur{Shariatpanahi}}
\author[1]{\fnm{José C.} \sur{García Alanis}}
\author[1]{\fnm{Mohammad-Ali} \sur{Nikouei Mahani}}
\author[3]{\fnm{Stefan G.} \sur{Hofmann}}
\author[4]{\fnm{Tim} \sur{Hahn}}
\author*[1,2,5]{\fnm{Hamidreza} \sur{Jamalabadi}}\email{hamidreza.jamalabadi@uni-marburg.de}

\affil[1]{\orgdiv{Department of Psychiatry and Psychotherapy}, \orgname{University of Marburg}, \country{Germany}}
\affil[2]{\orgdiv{Center for Mind, Brain, and Behavior (CMBB)}, \orgname{University of Marburg}, \country{Germany}}
\affil[3]{\orgdiv{Department of Psychology}, \orgname{University of Marburg}, \country{Germany}}
\affil[4]{\orgdiv{Institute for Translational Psychiatry}, \orgname{University of Münster}, \country{Germany}}
\affil[5]{\orgdiv{Faculty of Medicine}, \orgname{University of British Columbia}, \country{Canada}\medskip}


\abstract{In neuropsychiatry, the primary goal is often not only to decode brain activity but to change it, for example to lessen a negative affective bias or an overly salient memory. Motivated by control theory, we develop an AI-driven neural-surrogate framework that proposes candidate representational changes and tests their predicted perceptual effects from snapshots of stimulus-evoked fMRI activity, without physical stimulation. The framework combines fMRI decoding, deep generative modeling, and constrained latent-space steering. Valence and memorability are used only as worked examples. Using more than 36{,}000 image--fMRI observations from four deeply sampled Natural Scenes Dataset participants, subject-specific models recovered coarse generative structure from visually responsive cortex (two-way identification, $0.79$--$0.88$; chance, $0.5$). Graded perturbations were reconstructed as images and evaluated with automated scorers and human ratings from 7{,}200 trials by 18 participants. In the primary VDVAE model, valence shifted from $-0.61$ to $+1.03$~SD and memorability from $-1.34$ to $+1.45$~SD; a later Versatile Diffusion refinement reduced or altered these effects. Across five perturbation levels, human valence ratings moved in the predicted direction under the linear time-correction model (mean slope, $0.038$~SD per unit of $\alpha$; 95\% CI, $0.003$--$0.074$; positive in 16 of 18 participants). Perceived memorability did not change reliably. Baseline agreement with the automated assessor was suggestive for valence ($r = 0.30$) and weak for memorability ($r = 0.10$). Extreme perturbations drifted from the original stimulus, so intended change must be weighed against loss of fidelity. These findings provide a falsifiable upstream method for designing and behaviorally testing candidate representational targets for future neuromodulation in psychiatry, while marking the limits of the present static approximation.
}


\maketitle

\section{Introduction}\label{sec1}

Can we systematically design candidate perturbations of neural representations that may ultimately support the modulation of cognitive-affective properties relevant to mental health and neuropsychiatric symptoms, such as perceived valence \citep{barrett2006valence} and mnemonic salience \citep{rust2020understanding}? Doing so requires models that specify how a neural representation should change to produce a desired cognitive-affective outcome. This problem lies at the convergence of cognitive neuroscience, artificial intelligence (AI), and control theory \citep{kheirkhah2026re}, with important implications for biomedical engineering and psychiatric treatment \citep{gaziv2023strong,shanechi2019brain,hahn2023towards}. Clinically, persistent negative valence biases are prominent in Major Depressive Disorder, whereas overly salient or intrusive memories are central to conditions such as Post-Traumatic Stress Disorder \citep{cushing2023generative,hamilton2008neural,lemoult2019depression}. If neuromodulation could causally alter these representations, it could offer therapeutic value \citep{zhao2025transcutaneous,hofmann2025network}. At the most general level, this is a dynamical optimal-control problem: stimulation parameters act as control inputs, neural activity constitutes the evolving system state, and the objective is defined at the level of subjective or behavioral outcomes \citep{kirk2004optimal,lydon2021modeling,kheirkhah2026re}. Solving that complete problem would require a causal model linking stimulation, neural dynamics, and cognition.

The present study addresses the upstream target-identification step of this broader control problem. Before selecting a physical stimulation input or optimizing a neural trajectory, one must specify a candidate representational change associated with the desired outcome and determine whether its predicted consequences are behaviorally meaningful. We formulate this step as a static, local inverse-design problem. It contains neither a stimulation-to-brain forward model nor neural dynamics and is therefore a simplified component of a future control system, not a demonstration of closed-loop or dynamical neural control. Ashby's Good Regulator Theorem \citep{conant1970every,haimerl2025time} motivates this model-based progression from prediction toward intervention design. A predictive association indicates what variables covary; intervention design additionally requires a model that can generate explicit perturbation hypotheses and predict their consequences. Here, the theorem motivates asking whether a predictive brain--cognition model preserves sufficient task-relevant structure to propose candidate changes that can be tested before physical intervention; it does not imply that the present surrogate is already a controller or a causal model of stimulation.

\begin{figure}[t]
  \centering
  \includegraphics[width=0.99\textwidth]{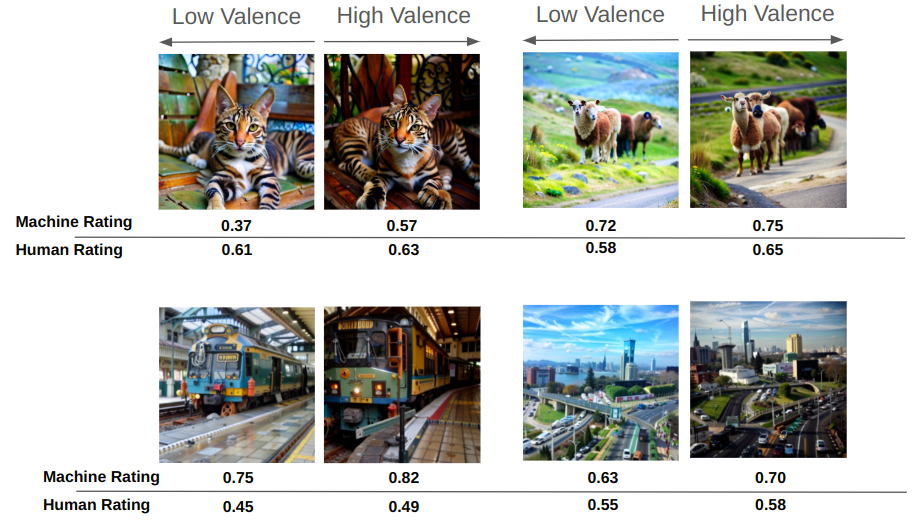}
\caption{\textbf{Conceptual Illustration of Surrogate-Guided In Silico Cognitive-Affective Neuromodulation Design.}
Reconstructed visual stimuli illustrate the predicted consequences of surrogate-guided representational perturbations targeting emotional valence.
Numerical scores below each image summarize automated valence estimates from EmoNet \citep{goetschalckx2019ganalyze} and independent human valence ratings; they do not represent cognitive changes directly induced through brain stimulation. Warmer hues and pleasant scenes are associated with positive-direction perturbations, whereas cooler and more aversive content is associated with negative-direction perturbations. These examples are illustrative and should not be used to infer effect size or consistency; baseline human--assessor correspondence and perturbation effects are evaluated in Sections~\ref{sec:baseline-agreement} and~\ref{effect_of_alpha}, respectively. 
Additional examples are available at \url{https://www.neuralsurrogate.com/}.}
\label{fig:Figure_1}
\end{figure}

Specifically, we construct subject-specific AI-driven neural surrogates that link stimulus-evoked fMRI activity to generative latent representations and cognitive-affective attributes. Under a local linear approximation and an explicit norm constraint, the resulting optimization problem admits a closed-form first-order solution; in practice, we use empirically estimated population-level steering directions as a tractable implementation, without assuming that they approximate each image's locally optimal gradient. We apply these perturbations across graded strengths, reconstruct their predicted perceptual consequences as images, and evaluate the resulting changes using model-based assessors and independent human ratings. The resulting latent perturbations and surrogate-predicted activity-pattern changes are treated as candidate representational targets for future neuromodulation. Translating them into implementable stimulation parameters will require a subsequent AI-based or biophysical forward model relating stimulation inputs to induced electric fields and propagated neural responses, potentially combining surrogate models of effective connectivity with multichannel stimulation optimization \citep{luo2025mapping,ruffini2014optimization,karimi2025precision}. Accordingly, the contribution of the present work is to generate and systematically test upstream representational targets, not the stimulation parameters needed to induce them (Figure~\ref{fig:Figure_1}).

\begin{figure}[t]
  \centering
  \includegraphics[width=0.99\textwidth]{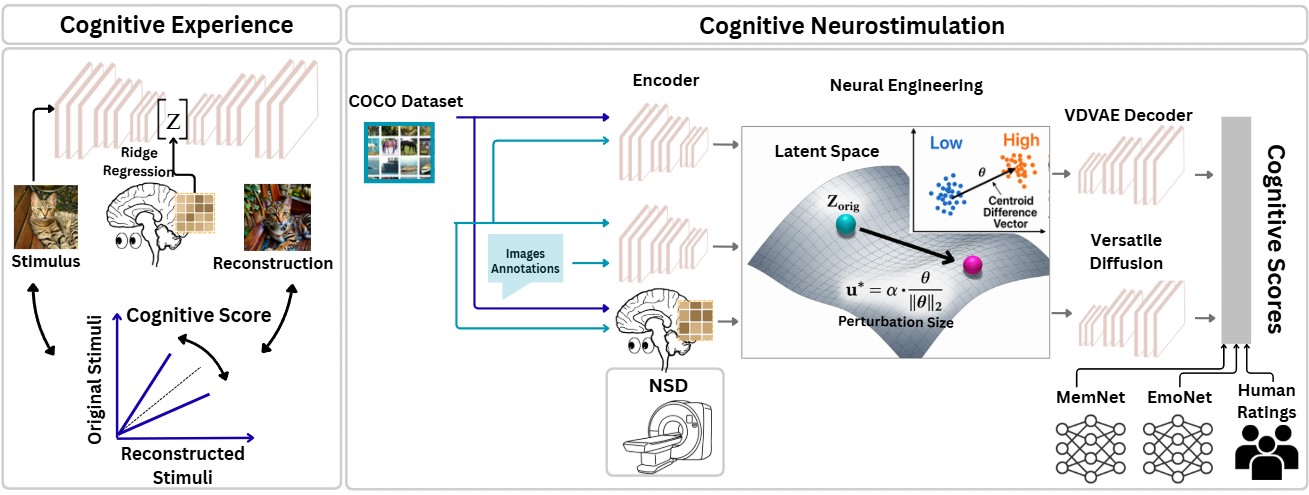}
 \caption{\textbf{Framework for AI-Driven In Silico Perturbation Design.}
  \textit{Left:} Subject-specific ridge regression maps stimulus-evoked 7T fMRI
  patterns from the Natural Scenes Dataset \cite{allen2022massive} into the
  latent space of a pretrained VDVAE \cite{child2020very}, enabling   reconstruction of the viewed stimuli. Valence and memorability are quantified   using EmoNet \cite{goetschalckx2019ganalyze} and MemNet
  \cite{khosla2015understanding}. \textit{Right:} Empirical attribute directions are applied at graded strengths to fMRI-decoded latent representations. The perturbed representations are reconstructed using a
  CLIP-conditioned diffusion pipeline   \cite{radford2021learning,ozcelik2023natural} and evaluated using model-based   scores, independent human ratings, and image-fidelity measures. The fMRI-space
  projection visualizes candidate activity-pattern directions under the learned   decoder; the pipeline does not derive physical stimulation parameters.}
\label{fig:surrogate-framework}
\end{figure}

We operationalize the framework in three stages, summarized in Figure~\ref{fig:surrogate-framework}. First, using 7T fMRI data from four deeply sampled Natural Scenes Dataset participants, each with more than $9{,}000$ image--fMRI observations \citep{allen2022massive}, we learn subject-specific mappings from stimulus-evoked activity in visually responsive cortex to the latent space of a pretrained deep autoencoder. Second, we derive a closed-form first-order perturbation solution under a local linear approximation and use empirical population-level steering directions to implement the valence and memorability perturbation series. These directions are applied at graded strengths and reconstructed as images using a CLIP-conditioned diffusion pipeline \citep{radford2021learning,ozcelik2023natural}, making their predicted consequences directly inspectable in stimulus space (theoretical formulation: Section~\ref{sec:control}; empirical implementation: Section~\ref{sec:perturbation-implementation}). Third, we evaluate the resulting dose--response patterns and reconstruction fidelity using model-based assessors and independent human judgments. Valence and memorability are quantified using EmoNet and MemNet \citep{goetschalckx2019ganalyze,khosla2015understanding} and evaluated against 7,200 ratings per attribute from 18 behavioral participants. Together, these analyses ask four linked questions: whether the surrogate preserves attribute-related structure at baseline; whether graded perturbations produce target-consistent changes; whether automated changes survive independent human judgment; and how far the representation can move before fidelity deteriorates. Baseline human--assessor correspondence was numerically stronger for valence than for perceived memorability, although the valence association was uncertain. Across the full perturbation range, human ratings support an overall valence direction under the primary correction model but no corresponding perceived-memorability effect. Versatile Diffusion attenuates or alters the initial VDVAE response, and extreme perturbations increasingly drift from the original stimulus. The pipeline therefore reveals a bounded regime in which intended change must be weighed against loss of fidelity.

\section{Theoretical Framework for In Silico Perturbation Design}
\label{sec:control}

To formalize this upstream target-identification step, we first distinguish it from the complete control-theoretic problem. A complete framework would connect a desired cognitive outcome to a target neural trajectory and then determine the stimulation inputs required to produce that trajectory, including state dynamics, time-varying inputs, feedback, and hardware constraints. The present study deliberately isolates a simpler preceding component: identifying candidate representational changes associated with a desired cognitive-affective shift. Because the available Natural Scenes Dataset measurements are trial-level summaries of stimulus-evoked BOLD activity, we formulate this component as a static, local inverse-design problem rather than as a dynamical optimal-control problem.

Previous work indicates that cognitive-affective attributes such as valence and memorability can be predicted from visual content and that artificial visual representations share measurable structure with stimulus-evoked cortical activity \citep{conwell2021perceptual,conwell2023controlled,muttenthaler2025aligning,pmlr-v285-rothermel24a}. Generative latent spaces also contain directions associated with graded changes in these attributes \citep{goetschalckx2019ganalyze,younesi2022controlling,gaziv2023strong}. These observations motivate linking time-aggregated fMRI response patterns to a generative latent space in which candidate perturbations can be defined and their predicted consequences reconstructed. In this study, the framework targets score-relevant information recoverable from visually responsive fMRI patterns; it does not assume that these patterns constitute the complete neural implementation of valence or memorability.

\subsection{Problem Formulation}
\label{sec:problem-formulation}

Let $\mathbf{z}\in\mathbb{R}^n$ denote a latent representation in a
generative model, and let $D$ denote the corresponding image decoder. We define
$f(\mathbf{z})$ as a scalar cognitive-affective objective associated with the
decoded image. In the present implementation,
\begin{equation}
    f(\mathbf{z}) = (A\circ D)(\mathbf{z}),
    \label{eq:latent-score}
\end{equation}
where $A$ is a pretrained assessor of emotional valence or memorability. Human
ratings are not used to construct the perturbation directions; they provide an
independent evaluation of their perceptual consequences.

Given an fMRI-decoded baseline representation $\mathbf{z}_{\mathrm{org}}$, we
seek a perturbation $\mathbf{u}\in\mathbb{R}^n$ that maximizes the predicted
attribute score within a fixed perturbation budget:
\begin{equation}
    \max_{\|\mathbf{u}\|_2\leq\varepsilon}
    f(\mathbf{z}_{\mathrm{org}}+\mathbf{u}),
    \label{eq:latent-obj}
\end{equation}
where $\varepsilon$ bounds the displacement from the baseline representation. 

This constraint limits movement in latent space but does not guarantee perceptual fidelity, which is therefore evaluated empirically from the reconstructed images. Equation~\ref{eq:latent-obj} describes the static representational target-design component of the broader neuromodulation problem. It is a constrained optimization problem, but it is not by itself a complete optimal-control formulation: it contains no state trajectory, feedback mechanism, time-varying stimulation input, or actuator model. This simplification is appropriate for the present analysis because each Natural Scenes Dataset observation is represented by a trial-level beta pattern summarizing stimulus-evoked BOLD activity \citep{allen2022massive}. The relation between this static component and a future dynamical control formulation is considered in Section~\ref{discussion}.

\subsection{Local First-Order Solution}

For a differentiable objective and sufficiently small $\varepsilon$, we
approximate $f$ around $\mathbf{z}_{\mathrm{org}}$ using a first-order Taylor
expansion:
\begin{equation}
    f(\mathbf{z}_{\mathrm{org}}+\mathbf{u})
    \approx
    f(\mathbf{z}_{\mathrm{org}})
    +\nabla f(\mathbf{z}_{\mathrm{org}})^{\top}\mathbf{u}.
    \label{eq:taylor-approx}
\end{equation}
Assuming a non-zero gradient, the perturbation that maximizes this linearized
objective under the $\ell_2$-norm constraint follows from the
Cauchy--Schwarz inequality:
\begin{equation}
    \mathbf{u}_{\mathrm{lin}}^{*}
    =
    \varepsilon
    \frac{\nabla f(\mathbf{z}_{\mathrm{org}})}
         {\|\nabla f(\mathbf{z}_{\mathrm{org}})\|_2}.
    \label{eq:latent-opt}
\end{equation}
This solution is locally optimal for the linearized objective: among perturbations within the specified norm budget, it produces the largest predicted first-order increase in $f$. It does not establish global optimality
for the nonlinear assessor--decoder pipeline or guarantee perceptual validity at larger perturbation magnitudes. The practical consequences of these limitations are evaluated empirically in the subsequent analyses.

\subsection{Empirical Population-Level Steering Direction}

The exact sample-specific gradient may be unavailable when the assessor or decoder is treated as a black box. A shared direction estimated from scored examples also provides a transparent and uniformly applicable steering rule.
We therefore estimate an empirical attribute axis from the difference between the centroids of high- and low-scoring latent representations
\citep{pmlr-v285-rothermel24a,goetschalckx2019ganalyze}:
\begin{equation}
    \boldsymbol{\theta}
    =
    \bar{\mathbf{z}}_{\mathrm{high}}
    -
    \bar{\mathbf{z}}_{\mathrm{low}},
    \label{eq:theta-def}
\end{equation}
where
\begin{equation}
    \bar{\mathbf{z}}_{\mathrm{high}}
    =
    \frac{1}{|\mathcal{H}|}
    \sum_{\mathbf{z}\in\mathcal{H}}\mathbf{z},
    \qquad
    \bar{\mathbf{z}}_{\mathrm{low}}
    =
    \frac{1}{|\mathcal{L}|}
    \sum_{\mathbf{z}\in\mathcal{L}}\mathbf{z}.
    \label{eq:centroids}
\end{equation}
Here, $\mathcal{H}$ and $\mathcal{L}$ contain the latent representations associated with the top and bottom quartiles of the training scores, respectively.

If the objective varies smoothly and approximately monotonically along the relevant region of the latent manifold, $\boldsymbol{\theta}$ summarizes a population-level direction separating low- and high-scoring examples. Held-out latent projections tracked the corresponding assessor scores for valence ($r=0.812$) and memorability ($r=0.834$; Appendix~\ref{app:theta_validation}). This validates a shared attribute axis, not an exact sample-specific gradient. The complementary exact-gradient benchmark in Appendix~\ref{sec:exact_gradient_bound} directly examines that distinction and the limits of single-direction steering.

The perturbations used to generate the experimental images are implemented as
\begin{equation}
\mathbf{u}_{\theta}
=
\alpha
\frac{\boldsymbol{\theta}}
{\|\boldsymbol{\theta}\|_2},
\label{eq:empirical-update}
\end{equation}
where the sign of $\alpha$ determines the perturbation direction and its magnitude determines perturbation strength. Equation~\ref{eq:latent-opt} gives the locally optimal solution for the linearized objective when the exact sample-specific gradient is available. Equation~\ref{eq:empirical-update}, by contrast, is the empirical population-level steering rule implemented in the present experiments. It provides a transparent, black-box-compatible population-level steering direction and generalizes established linear latent-editing approaches \citep{goetschalckx2019ganalyze}; it is not an exact optimizer of the full nonlinear pipeline. Claims of local optimality therefore apply to the first-order solution in Equation~\ref{eq:latent-opt}, not automatically to the empirical direction $\boldsymbol{\theta}$. The main experimental images were generated by applying Equation~\ref{eq:empirical-update} directly to latent representations decoded from fMRI.

\subsection{Candidate Directions in fMRI Response Space}

To relate the latent perturbations to the measured brain responses, we learn a subject-specific linear ridge decoder
\begin{equation}
    \mathbf{z}=R\mathbf{x},
    \label{eq:fmri-to-latent}
\end{equation}
where $\mathbf{x}\in\mathbb{R}^d$ is a voxelwise fMRI beta pattern and $R\in\mathbb{R}^{n\times d}$ maps that pattern into the generative latent
space \citep{ozcelik2023natural}. This is a statistical decoder from fMRI measurements to latent representations; it is not a forward model of how a
stimulus or stimulation input generates a neural response.

The surrogate objective in fMRI response space is
\begin{equation}
    g(\mathbf{x})
    =
    f(R\mathbf{x})
    =
    (A\circ D\circ R)(\mathbf{x}).
    \label{eq:fmri-score}
\end{equation}
The corresponding local optimization problem is
\begin{equation}
    \max_{\|\mathbf{u}_x\|_2\leq\varepsilon_x}
    g(\mathbf{x}_{\mathrm{org}}+\mathbf{u}_x).
    \label{eq:fmri-obj}
\end{equation}
Because $R$ is linear, the gradient of the surrogate objective with respect to
the measured fMRI pattern is
\begin{equation}
    \nabla_{\mathbf{x}}g(\mathbf{x})
    =
    R^{\top}\nabla_{\mathbf{z}}f(R\mathbf{x}).
    \label{eq:fmri-gradient}
\end{equation}
The locally optimal first-order direction under an $\ell_2$ constraint is
therefore
\begin{equation}
    \mathbf{u}_{x,\mathrm{lin}}^{*}
    =
    \varepsilon_x
    \frac{R^{\top}\nabla_{\mathbf{z}}f(\mathbf{z}_{\mathrm{org}})}
         {\|R^{\top}\nabla_{\mathbf{z}}f(\mathbf{z}_{\mathrm{org}})\|_2}.
    \label{eq:fmri-opt}
\end{equation}
Substituting the empirical attribute direction gives the candidate fMRI-pattern
direction
\begin{equation}
    \mathbf{u}_{x,\theta}
    \propto
    R^{\top}\boldsymbol{\theta}.
    \label{eq:fmri-empirical}
\end{equation}

Here, $R^{\top}$ acts as the adjoint of the learned decoder: it pulls the latent attribute direction back into the measured fMRI feature space. It is not an inverse causal model, a model of neural dynamics, or a stimulation control law, and it does not imply that the resulting pattern can be physically induced. Rather, $R^{\top}\boldsymbol{\theta}$ expresses a candidate fMRI-pattern sensitivity direction associated with movement along the empirical latent attribute axis under the fitted surrogate. In this study, this formulation is used to express and visualize candidate activity-pattern changes; implementation details for the cortical maps are provided in Appendix~\ref{app:cortical_map_computation}. The reconstructed images and human evaluations are based on the direct latent-space perturbations in Equation~\ref{eq:empirical-update}, not on physical or simulated application of $R^{\top}\boldsymbol{\theta}$ to the brain.

\section{Materials and Methods}
\label{sec:methods}

Following the theoretical formulation, we describe the empirical pipeline in the same order: construction of the neural surrogate, implementation and reconstruction of the perturbations, automated assessment, and independent human evaluation.

\subsection{Neural Surrogate and Image Reconstruction}
\label{subsec:neural-surrogate}

\subsubsection{Natural Scenes Dataset and fMRI preprocessing}
\label{sec:reconstruction}

We used the Natural Scenes Dataset (NSD), a large-scale 7~T fMRI dataset in
which eight participants viewed up to 10,000 natural images drawn from the
Microsoft COCO dataset \citep{allen2022massive,lin2014microsoft}. The present
analyses used data from four NSD participants and followed the reconstruction
pipeline of \citet{ozcelik2023natural}. Visually responsive voxels were selected
using each participant's NSD-general region of interest. The available
single-trial beta estimates were aggregated by image, yielding 8,859 training
samples and 982 held-out test samples per participant. The held-out images were
not used to fit the subject-specific mappings.

Before regression, fMRI responses were divided by 300 and standardized using
the mean and standard deviation estimated from the training data; the same
parameters were then applied to the test data. All mappings were fitted and
evaluated separately for each participant because voxelwise fMRI features are
not directly aligned across individuals. No cross-participant anatomical or
functional alignment was performed.

\subsubsection{Subject-specific fMRI-to-VDVAE mapping}
\label{sec:fmri-vdvae-method}

Each stimulus image was encoded with a pretrained Very Deep Variational
Autoencoder (VDVAE) \citep{child2020very}. We used the frozen ImageNet-64 VDVAE
checkpoint, which represents each image by a concatenated latent vector
$\mathbf{z}\in\mathbb{R}^{n}$ with $n=91{,}168$. These image-derived latent
vectors served as targets for learning the statistical mapping from fMRI
activity to the generative latent space.

Following prior fMRI reconstruction work
\citep{ozcelik2023natural,pmlr-v285-rothermel24a}, we fitted a subject-specific
ridge decoder. Let $X_{\mathrm{train}}^{(s)}\in\mathbb{R}^{N\times d_s}$ denote
the training fMRI patterns for subject $s$ and
$Z_{\mathrm{train}}\in\mathbb{R}^{N\times n}$ the corresponding VDVAE latent
vectors. Consistent with Equation~\ref{eq:fmri-to-latent}, the decoder matrix
$R^{(s)}\in\mathbb{R}^{n\times d_s}$ was estimated as
\begin{equation}
    R^{(s)}
    =
    \arg\min_R
    \left\|Z_{\mathrm{train}}-X_{\mathrm{train}}^{(s)}R^{\top}\right\|_F^2
    +\lambda\left\|R\right\|_F^2.
    \label{eq:ridge-method}
\end{equation}
An intercept was included, and the VDVAE decoder used a ridge penalty of
$\lambda=50{,}000$. The trained model produced held-out latent estimates
\begin{equation}
    \widehat{Z}_{\mathrm{test}}^{(s)}
    =
    X_{\mathrm{test}}^{(s)}R^{(s)\top}+\mathbf{b}^{(s)}.
    \label{eq:latent-prediction-method}
\end{equation}
The VDVAE was not retrained or fine-tuned. Its frozen decoder transformed the
predicted latent vectors into coarse image reconstructions, which were resized
to $512\times512$ pixels for subsequent processing. Additional architectural
details are provided in Appendix~\ref{vae recon}.

The learned mapping is a statistical decoder from measured fMRI patterns to
image latents. It is not a causal forward model of neural activity and is not
used here to infer physically implementable stimulation inputs.

\subsubsection{Construction of the latent perturbation series}
\label{sec:perturbation-implementation}

Candidate perturbations were constructed in the VDVAE latent space using the
empirical population-level direction defined in
Equation~\ref{eq:theta-def}. For each target domain, the training images were
scored with the corresponding frozen assessor. The latent centroids of the top
and bottom score quartiles were computed, and their difference defined the
attribute direction $\boldsymbol{\theta}$. For each fMRI-decoded test latent
$\widehat{\mathbf{z}}_{0}^{(s)}$, the perturbation series was generated as
\begin{equation}
    \widehat{\mathbf{z}}_{\alpha}^{(s)}
    =
    \widehat{\mathbf{z}}_{0}^{(s)}
    +
    \alpha\frac{\boldsymbol{\theta}}
    {\|\boldsymbol{\theta}\|_2},
    \qquad
    \alpha\in\{-4,-2,0,2,4\}.
    \label{eq:alpha-series-method}
\end{equation}
Negative and positive values of $\alpha$ specify the intended directions of
decreasing and increasing attribute scores, respectively. Images used in the main
automated and human evaluations were generated by applying
Equation~\ref{eq:alpha-series-method} directly in latent space. The
fMRI-pattern direction in Equation~\ref{eq:fmri-empirical} was used only as a
surrogate-derived sensitivity representation; it was not applied to measured
brain activity and was not converted into stimulation parameters.

\subsubsection{Versatile Diffusion refinement}
\label{sec:diffusion-refinement}

We used a frozen Versatile Diffusion model to refine the coarse VDVAE outputs
into higher-resolution RGB images \citep{xu2023versatile}. This stage follows
the low-level/high-level reconstruction strategy of
\citet{ozcelik2023natural}: the VDVAE output provides the image scaffold, while
CLIP text and vision embeddings provide semantic conditioning
\citep{radford2021learning}.

The conditioning source differed between the two uses of this stage. Baseline
brain reconstructions were conditioned on CLIP text and vision embeddings
predicted from fMRI, preserving an end-to-end fMRI reconstruction pathway. For
the perturbed $\alpha$-series, the diffusion model was conditioned on CLIP
embeddings obtained from the corresponding original test image and caption.
Holding this semantic conditioning fixed reduced variation unrelated to the
VDVAE perturbation, but it also means that the refined $\alpha$-series images
are not end-to-end fMRI-only reconstructions. We identify this distinction
explicitly when presenting the results. Full model settings and reconstruction
parameters are reported in Appendix~\ref{app:reconstruction_implementation}, Sections~\ref{clip reconstruction 1} and~\ref{clip reconstruction 3}.

Reconstruction fidelity at each perturbation level was characterized relative
to the corresponding unperturbed reconstruction $I_0$ using two complementary
descriptive measures: pixel correlation,
$\operatorname{PixCorr}(I_{\alpha},I_0)$, and CLIP image-embedding similarity,
\begin{equation}
    s_{\mathrm{CLIP}}(I_{\alpha},I_0)
    =
    \cos\!\left(
      \operatorname{CLIP}(I_{\alpha}),
      \operatorname{CLIP}(I_0)
    \right),
    \label{eq:clip-similarity-method}
\end{equation}
computed with the CLIP ViT-L/14 vision encoder. These measures were used to
describe response--fidelity trade-offs and the operating range of the local
perturbation approximation. All five perturbation levels were included in the full-range analyses, and no image was excluded. The $\alpha=\pm4$ levels were additionally treated as boundary conditions when evaluating fidelity and the limits of the local approximation (Section~\ref{trade off}). A separate image-quality sensitivity analysis using PixCorr and structural similarity (SSIM) is reported in Appendix~\ref{app:qc_sensitivity}; these thresholds did not define exclusions for the main full-range results.

\subsection{Automated Assessment of Image Attributes}
\label{sec:assess}

We quantified the two target attributes with pretrained assessor networks used
in frozen, inference-only mode. MemNet estimates image memorability and was
trained on the LaMem dataset of more than 60,000 images with human memory scores
\citep{khosla2015understanding}. Its output ranges from 0 to 1, with larger
values indicating greater predicted memorability. The EmoNet assessor described in GANalyze provides an
image-derived estimate of emotional valence, with larger values indicating more
positive predicted valence \citep{goetschalckx2019ganalyze}. These scores are
model-based proxies for the selected image attributes and are not direct
measurements of participants' cognitive or affective states.

Assessor scores were standardized within each fMRI subject and reconstruction
stream before group-level model summaries were computed. Human ratings were not
used to estimate the perturbation directions. Instead, they served as an
independent test of whether the score changes predicted by the in silico
pipeline were detectable in judgments of the reconstructed images.

\subsection{Independent Human Evaluation of Reconstructed Stimuli}
\label{human experiment method}

\subsubsection{Ethics and participants}

The online behavioral study was reviewed and approved by the Ethics Committee
of the University of Marburg. All participants provided written informed
consent and received compensation consistent with local standards. The study
measured ratings of model-generated images only: participants did not undergo
fMRI or neurostimulation, did not contribute data to model training, and were
not exposed to a physical neuromodulation protocol.

Twenty participants were recruited; 18 completed the experiment and were
included in the final analysis (10\% dropout). Participants were 21--39 years
old: 12 (66.7\%) were aged 21--29 and 6 (33.3\%) were aged 30--39. Ten
participants (55.6\%) identified as female and eight (44.4\%) as male. Ages are
reported in ranges to comply with the anonymity requirements of the local
ethics committee. Further demographic information is summarized in
Table~\ref{tab:demographics}.

\begin{table}[t]
\centering
\caption{Participant characteristics for the independent behavioral evaluation
($N=18$).}
\label{tab:demographics}
\begin{tabular}{llrr}
\toprule
Variable & Category & Count & Percent \\
\midrule
Age group & 21--29 & 12 & 66.7 \\
          & 30--39 & 6 & 33.3 \\
Gender & Female & 10 & 55.6 \\
       & Male & 8 & 44.4 \\
Mental-health history & No, never & 12 & 66.7 \\
                      & Yes, in the past & 2 & 11.1 \\
                      & Yes, currently in treatment & 3 & 16.7 \\
                      & No, but seeking support & 1 & 5.6 \\
Education & High school & 5 & 27.8 \\
          & MA/M.Sc. & 5 & 27.8 \\
          & BA/B.Sc. & 3 & 16.7 \\
          & PhD & 3 & 16.7 \\
          & Other & 2 & 11.1 \\
\bottomrule
\end{tabular}
\end{table}

\subsubsection{Procedure and rating task}

The experiment was implemented in PsychoPy and hosted online through Pavlovia
\citep{peirce2019psychopy2}. Participants first completed the Patient Health
Questionnaire (PHQ-9) \citep{kroenke2001phq}. Under the study's prespecified
safety rule, participants with scores greater than 10 received an automated
recommendation to seek professional support. A brief sociodemographic survey
and tutorial then introduced the three rating dimensions---arousal, valence,
and perceived memorability---using continuous visual analogue scales from 0 to
100. Practice trials familiarized participants with the task.

The main task contained 400 image presentations arranged in ten blocks of 40
trials and lasted approximately 140 minutes, including self-paced breaks. The
ten conditions comprised original NSD images, unperturbed fMRI-based
reconstructions ($\alpha=0$), and reconstructions perturbed toward lower or
higher valence or memorability at $\alpha\in\{-4,-2,2,4\}$. Thus, the
behavioral experiment evaluated perceptual judgments of reconstructed stimuli;
it did not test whether the candidate neural patterns could be induced in the
brain.

Before each block, participants viewed the same standardized demonstration
screen. The examples were natural images from the Open Affective Standardized
Image Set (OASIS), selected using its normative valence and arousal ratings
\citep{kurdi2017introducing}. These images were neither NSD stimuli nor model-generated
images and were excluded from all analyses. Memorability was demonstrated
separately as a low-to-high dimension.

On each trial, an image was presented for 350~ms and was followed by sequential
ratings of arousal, valence, and perceived memorability. Arousal was collected
to preserve a consistent rating interface but was not a target of the
perturbation design and was not included in the primary analyses. Participants
were not informed about the model targets, perturbation strengths, or image
generation procedure until debriefing.

\subsubsection{Behavioral analysis}

Across the 18 included participants, the task yielded 7,200 image-presentation
trials and 7,200 ratings for each of the three dimensions. Each of the ten
conditions contributed 40 trials per participant (720 observations before
averaging). One condition contained 39 unique source images because one image
was presented twice; repeated observations of the same
participant--condition--image combination were averaged before analysis. Among
the 18 completers included in the analysis, no trial or image was excluded from
the primary analysis.

For valence and memorability, all five perturbation levels ($\alpha\in\{-4,-2,0,+2,+4\}$) were used to summarize full-range directional trends. The $\alpha=\pm4$ levels were additionally treated as boundary conditions when evaluating fidelity and breakdown of the local approximation.

Because all participants completed the same fixed block order (specified in Appendix~\ref{app:behavioral_sensitivity}), raw 0--100 ratings were adjusted using leave-one-participant-out, assessor-neutral linear correction for source-image composition, within-block position, and between-block time. For each held-out participant, nuisance coefficients were estimated from the other 17 participants using the unperturbed trials and perturbed images with absolute final-stage assessor change no larger than 0.50 baseline SD. The adjusted condition mean was expressed relative to the adjusted $\alpha=0$ mean and divided by that participant's $\alpha=0$ rating standard deviation. We then estimated each participant's linear slope across all five levels. Restricted-range slopes ($\alpha=-2$ to $+2$ and $-2$ to $+4$) were retained as sensitivity analyses, not as separate primary outcomes. We report mean slopes, unadjusted 95\% confidence intervals across participants, positive-slope counts, and strictly increasing curves. Directional consistency was evaluated using an exact binomial test against a positive-slope probability of 0.5 and an exact sign-flip test of the mean slope; the reported $p$ values retain Holm adjustment across the six target-by-range comparisons, including the sensitivity ranges.

Inter-rater reliability was quantified separately for each condition using the two-way random-effects single-rater ICC(2,1) and average-rater ICC(2,$k$), supplemented by split-half reliability \citep{shrout1979intraclass}. Baseline human--assessor agreement was calculated across the 40 source-image indices in the unperturbed behavioral block: ratings were averaged across participants within image and correlated with the corresponding $\alpha=0$ Versatile Diffusion assessor scores, standardized within each fMRI subject and then averaged across the four fMRI subjects. The behavioral files did not record the fMRI-subject provenance of each displayed reconstruction, so this comparison is source-index matched rather than a verified match to the exact rendered image. It does not use the separate image-to-image reference scores. Correlation confidence intervals were calculated on the Fisher-$z$ scale, with unadjusted two-sided $p$ values. These correlations quantify correspondence between model-based and human judgments, not causal neuromodulation.

The correction rationale, equations, alternative specifications, and exact-source matched analyses are reported in Appendix~\ref{app:behavioral_sensitivity}; image-quality sensitivity is reported in Appendix~\ref{app:qc_sensitivity}. Figure~\ref{fig:experiment-concept} summarizes the behavioral procedure and its separation from any physical neuromodulation protocol.

\begin{figure}[t]
  \centering
  \includegraphics[width=0.99\textwidth]{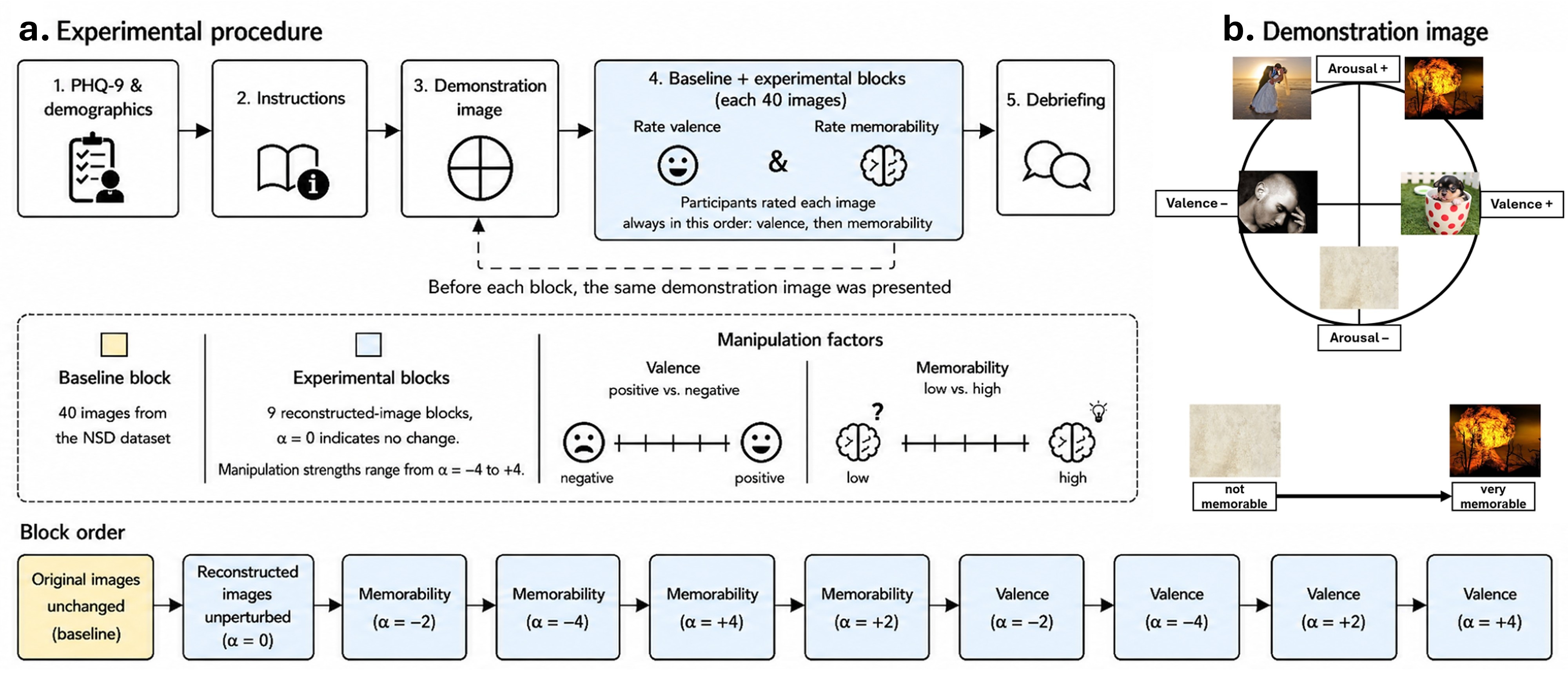}
  \caption{\textbf{Independent human evaluation of reconstructed stimuli.}
  \textbf{(a)} Participants completed the PHQ-9 and a demographic survey,
  followed by task instructions and practice. The main experiment contained
  one original-image block and nine reconstruction blocks, with 40 trials per
  block. The reconstruction conditions comprised the unperturbed baseline
  ($\alpha=0$) and valence- or memorability-targeted perturbations at
  $\alpha\in\{-4,-2,2,4\}$. Each image was shown for 350~ms and followed by
  ratings of arousal, valence, and perceived memorability. \textbf{(b)} Before
  each block, a fixed demonstration screen illustrated the valence--arousal
  space using OASIS images \citep{kurdi2017introducing}; memorability was demonstrated
  separately. Demonstration images were not drawn from NSD, were not generated
  by the surrogate pipeline, and were excluded from the analyses. Participants
  rated only the perceptual consequences of the in silico perturbations; no
  neurostimulation was administered.}
  \label{fig:experiment-concept}
\end{figure}

\section{Results}\label{subsec:results}

\subsection{Baseline Preservation and Human--Assessor Agreement}
\label{sec:baseline-agreement}

We first examined whether attribute-related score structure was preserved at the unperturbed baseline and whether the automated assessments corresponded to independent human judgments (Table~\ref{tab:baseline_agreement}).

Automated score preservation was strong after Versatile Diffusion reconstruction. The correlation between assessor scores at the unperturbed baseline and the corresponding image-to-image reconstructions was \(r=0.63\) for valence (95\% CI: [0.61, 0.65], \(p<0.001\)) and \(r=0.77\) for memorability (95\% CI: [0.75, 0.79], \(p<0.001\)). Thus, the final reconstruction stage largely preserved the relative ordering of images along both assessor-defined dimensions.

We next compared mean human ratings with source-matched, subject-averaged $\alpha=0$ Versatile Diffusion assessor scores for the 40 images in the unperturbed behavioral condition. Human valence ratings showed a positive but uncertain correspondence with EmoNet scores (\(r=0.30\), 95\% CI: [$-0.01$, 0.56], \(p=0.058\)). The association between perceived-memorability ratings and MemNet scores was weak (\(r=0.10\), 95\% CI: [$-0.22$, 0.40], \(p=0.551\)). Thus, the reconstruction pipeline preserved automated attribute structure for both targets, while correspondence with human judgments was suggestive for valence and weak for perceived memorability. Neither baseline association met the conventional two-sided 0.05 threshold; their numerical difference is descriptive, not a formal test of a difference between correlations.

\begin{table}[t]
\centering
\small
\caption{\textbf{Baseline score preservation and human--assessor agreement.}
Automated score preservation was quantified by correlating assessor scores at the unperturbed baseline with scores from the corresponding final-diffusion image-to-image reconstructions. Correlations were Fisher-\(z\) averaged across the four fMRI subjects, with $p$ values Holm-adjusted across the 20 preservation tests. Human--assessor agreement was calculated across the 40 source-image indices in the unperturbed behavioral condition by correlating mean human ratings with standardized $\alpha=0$ Versatile Diffusion assessor scores averaged across the four fMRI subjects, rather than with the separate image-to-image reference scores. This comparison is source-index matched; the fMRI-subject provenance of the displayed images was not recorded. Human--assessor confidence intervals use the Fisher-$z$ transformation, and their two-sided $p$ values are unadjusted.}
\label{tab:baseline_agreement}
\begin{tabular}{
  >{\raggedright\arraybackslash}p{2.2cm} 
  >{\raggedright\arraybackslash}p{2.6 cm} 
  >{\raggedright\arraybackslash}p{2.5cm} 
  >{\raggedright\arraybackslash}p{2.3cm} 
  >{\centering\arraybackslash}p{2.4cm}   
  >{\centering\arraybackslash}p{1.32cm}   
}
\toprule
\textbf{Property} &
\textbf{Comparison} &
\textbf{Evaluator} &
\textbf{\(n\)} &
\textbf{\(r\) [95\% CI]} &
\textbf{\(p\)} \\
\midrule
\multirow{2}{*}{\textbf{Valence}} 
& Score preservation & EmoNet & 4 fMRI subjects & 0.63 [0.61, 0.65] & \(<0.001\) \\
& Human--assessor agreement & Human ratings--EmoNet & 40 images & 0.30 [$-0.01$, 0.56] & 0.058 \\
\midrule
\multirow{2}{*}{\textbf{Memorability}} 
& Score preservation & MemNet & 4 fMRI subjects & 0.77 [0.75, 0.79] & \(<0.001\) \\
& Human--assessor agreement & Human ratings--MemNet & 40 images & 0.10 [$-0.22$, 0.40] & 0.551 \\
\bottomrule
\end{tabular}
\end{table}

\subsection{Quantifying Attribute-Score Changes Under Perturbation}
\label{effect_of_alpha}
\begin{figure}[H]
\centering
\includegraphics[width=0.99\textwidth]{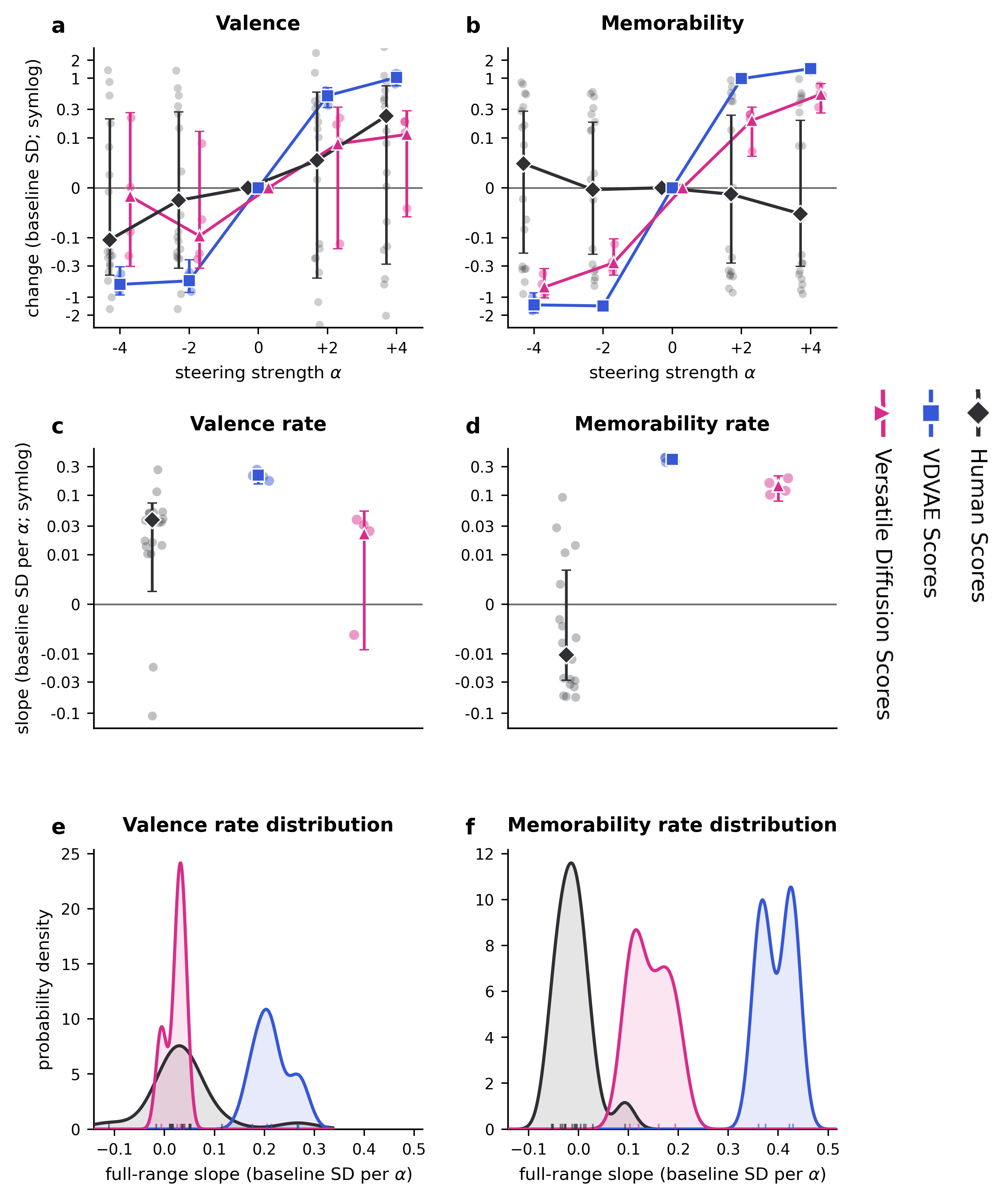}
\caption{\textbf{Attribute-score changes across latent perturbation strengths.}
\textbf{(a--b)} Changes relative to the unperturbed condition ($\alpha=0$), expressed in baseline standard-deviation units, for valence and memorability. Individual observations are shown as unconnected points; lines connect group means, and error bars denote 95\% confidence intervals. The symmetric-logarithmic axes have a linear region within $\pm0.10$ SD.
\textbf{(c--d)} Participant- and assessor-level slopes estimated across the full range from $\alpha=-4$ to $\alpha=+4$, in SD units per unit of $\alpha$. Error bars denote 95\% confidence intervals, and the symmetric-logarithmic axes have a linear region within $\pm0.01$ SD per unit of $\alpha$.
\textbf{(e--f)} Descriptive kernel-density estimates and rug plots of the same full-range slopes. These distributions are shown for visualization rather than inferential comparison, particularly because the automated estimates contain four fMRI subjects, compared with 18 human participants. Human estimates incorporate the leave-one-participant-out, assessor-neutral linear fatigue correction. The shared legend identifies Human Scores, VDVAE Scores, and Versatile Diffusion Scores.}
\label{fig:Figure_4}
\end{figure}

Having characterized baseline score preservation and human--assessor correspondence, we next examined whether movement along the empirical attribute directions produced systematic changes in valence and memorability. Perturbations were applied at strengths $\alpha \in \{-4,-2,0,+2,+4\}$, thereby testing both the direction and magnitude of cognitive-score changes across the VDVAE and Versatile Diffusion stages. Figure~\ref{fig:Figure_4} reports the change relative to the unperturbed reconstruction ($\alpha=0$), expressed in baseline standard-deviation units.

Three observations stand out. First, the VDVAE stage exhibited strong and largely graded modulation (Figure~\ref{fig:Figure_4}a--b). Mean valence changes increased strictly monotonically from $-0.61$ SD at $\alpha=-4$ to $+1.03$ SD at $\alpha=+4$. Memorability showed an equally clear directional separation, ranging from $-1.34$ SD at $\alpha=-4$ to $+1.45$ SD at $\alpha=+4$, although the two negative conditions displayed a small local reversal: the estimated reduction was slightly larger at $\alpha=-2$ than at $\alpha=-4$. Thus, the empirical directions produced systematic steering of both assessor-defined attributes at the VDVAE stage, even if the response was not perfectly monotonic at every adjacent memorability level.

Second, Versatile Diffusion altered the transmission of these perturbations. Memorability remained monotonic but was substantially attenuated, increasing from $-0.67$ SD at $\alpha=-4$ to $+0.54$ SD at $\alpha=+4$. Valence modulation, by contrast, was largely compressed: the estimated mean changes ranged only from $-0.10$ to $+0.11$ SD, and all corresponding confidence intervals included zero. The behavior at the negative extreme was particularly informative. Whereas VDVAE produced a pronounced negative-valence shift at $\alpha=-4$, the Versatile Diffusion estimate returned close to baseline ($-0.02$ SD). This indicates that the final generative stage does not transmit all latent perturbations proportionally and may regularize or overwrite strong manipulations, particularly at extreme values of $\alpha$.

Third, the human results differed across the two cognitive attributes. Single-rater agreement was low (ICC(2,1): 0.07--0.20 for valence and 0.09--0.12 for perceived memorability), whereas image ratings averaged over 18 raters were more reliable (ICC(2,$k$): 0.56--0.82 and 0.65--0.71, respectively; split-half reliability: 0.67--0.87). These reliability estimates describe agreement across raters; the slope analyses below retain individual participants as the units of analysis. For valence, corrected human ratings followed the expected overall direction, increasing from $-0.11$ SD at $\alpha=-4$ to $+0.23$ SD at $\alpha=+4$. The full-range participant-level slope was positive ($M=0.038$ SD per unit of $\alpha$, 95\% CI [0.003, 0.074]); sixteen of 18 participants had positive slopes (Holm-adjusted exact binomial $p=0.0039$), whereas the Holm-adjusted sign-flip test of the mean slope was borderline ($p=0.0527$). The participant signs were therefore predominantly positive, but evidence for a nonzero mean slope depended on the inferential summary. Nevertheless, only one participant displayed a strictly monotonic sequence across all five levels, and the confidence intervals for the individual $\alpha$-specific means included zero. The human evidence therefore supports an overall positive valence contrast under the linear correction model rather than a uniform monotonic dose--response at every perturbation level. This main estimate includes all five levels. Restricting the analysis to $\alpha=-2,0,+2$ gave a slope of 0.020 SD per unit of $\alpha$ (95\% CI [$-0.044$, 0.084]; 10 of 18 positive slopes), while the image-quality sensitivity estimates varied with the exclusion threshold (Appendix~\ref{app:qc_sensitivity}). These checks qualify the robustness of the full-range result rather than redefine its analysis range.

No corresponding effect was observed for perceived memorability. Corrected human mean changes remained close to zero across the perturbation range, from $+0.05$ SD at $\alpha=-4$ to $-0.05$ SD at $\alpha=+4$. The full-range slope was slightly negative but uncertain ($M=-0.011$ SD per unit of $\alpha$, 95\% CI [$-0.028$, 0.007]), and only 5 of 18 participants showed a positive slope. Thus, although the automated memorability assessor responded strongly to the perturbation, participants did not perceive a systematic change in memorability. Importantly, the behavioral task measured perceived memorability rather than subsequent memory performance; the implications of this distinction are considered further in the Discussion.

The participant- and assessor-level slopes in Figure~\ref{fig:Figure_4}c--d summarize these differences over the full perturbation range. For valence, the mean slopes were 0.216 SD per unit of $\alpha$ for VDVAE (95\% CI [0.155, 0.278]), 0.022 for Versatile Diffusion (95\% CI [$-0.009$, 0.054]), and 0.038 for human ratings. For memorability, the corresponding slopes were 0.397 for VDVAE (95\% CI [0.343, 0.452]), 0.145 for Versatile Diffusion (95\% CI [0.079, 0.210]), and $-0.011$ for human ratings. The descriptive distributions in Figure~\ref{fig:Figure_4}e--f show the same stage-dependent pattern: VDVAE produced the largest changes, Versatile Diffusion retained a weaker memorability effect but little valence modulation, and the human memorability slopes remained concentrated around zero.

Because the behavioral conditions were presented in a fixed block order, perturbation condition was confounded with between-block time. Human estimates were therefore obtained after leave-one-participant-out, assessor-neutral linear fatigue correction and should be interpreted conditional on that correction model. Appendix~\ref{app:behavioral_sensitivity}, particularly Figures~\ref{fig:appendix_behavioral_diagnostics}, \ref{fig:appendix_correction_sensitivity}, and~\ref{fig:appendix_exact_source}, reports the uncorrected results, alternative correction specifications, negative-control analyses, participant-level trajectories, and exact-source matched analyses. These analyses characterize the model dependence of the directional valence result while showing no robust evidence for perceived-memorability modulation. The effects of perturbation strength on image fidelity, including the degradation observed at extreme values of $\alpha$, are examined separately in Section~\ref{trade off}.

\subsection{Trade-offs Between Attribute-Score Modulation and Visual Fidelity}
\label{trade off}
\begin{figure}[H]
\centering
\includegraphics[width=0.99\textwidth]{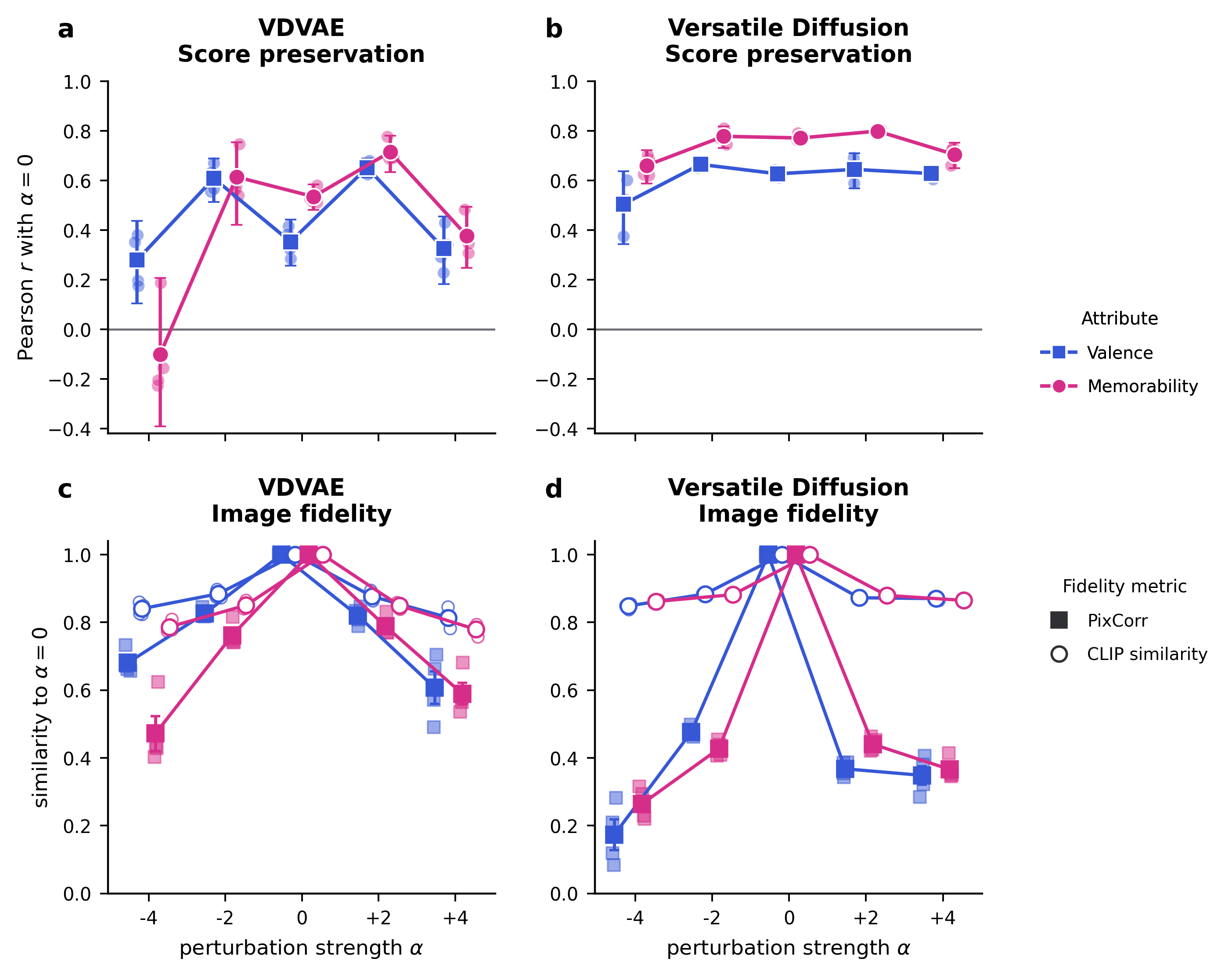}
\caption{\textbf{Trade-offs between attribute-score modulation and visual fidelity.}
\textbf{(a--b)} Pearson correlations between assessor scores at each perturbation level and the corresponding scores of the unperturbed reconstructions ($\alpha=0$), shown for the VDVAE and Versatile Diffusion stages. At $\alpha=0$, the correlations quantify score preservation between the unperturbed reconstruction and its image-to-image reconstruction.
\textbf{(c--d)} PixCorr and CLIP image-embedding similarity relative to the corresponding $\alpha=0$ reconstruction. Fidelity measures are oriented so that higher values indicate better preservation. Small, semi-transparent symbols show the four fMRI subjects and are not connected. Lines join only group means; large symbols and error bars show Fisher-$z$ mean correlations with 95\% confidence intervals in panels \textbf{a--b}, and means $\pm$ SEM in panels \textbf{c--d}. Color denotes the targeted attribute: valence (blue) or memorability (magenta). To improve readability, the valence and memorability series are slightly horizontally offset; this visual jitter has no numerical meaning. In the fidelity panels, filled squares denote PixCorr and open circles denote CLIP similarity.}

\label{fig:Figure_5}
\end{figure}

The changes in cognitive scores reported above must be interpreted together with the extent to which the perturbed images preserve the structure and visual content of the unperturbed reconstructions. We therefore quantified two complementary properties (Figure~\ref{fig:Figure_5}): assessor-score preservation, defined as the correlation between scores at each perturbation level and the corresponding scores at $\alpha=0$, and image fidelity, measured relative to $\alpha=0$ using PixCorr and CLIP image-embedding similarity.
At the VDVAE stage, valence-score preservation remained positive across all perturbation levels ($r=0.28$--$0.65$; Figure~\ref{fig:Figure_5}a). Memorability was also preserved at most levels, but broke down at the negative extreme: at $\alpha=-4$, the correlation with the unperturbed scores was $r=-0.10$ (95\% CI [$-0.39,0.21$]; Holm-adjusted $p=0.378$). Following Versatile Diffusion, score-preservation correlations were positive at every perturbation level for both valence ($r=0.50$--$0.66$) and memorability ($r=0.66$--$0.80$; Figure~\ref{fig:Figure_5}b). This preservation of relative image ordering does not imply successful transmission of the intended mean shift. In particular, Versatile Diffusion preserved valence rank ordering while substantially attenuating the mean valence modulation observed at the VDVAE stage (Figure~\ref{fig:Figure_4}).
Image fidelity decreased as the perturbations moved away from $\alpha=0$ (Figure~\ref{fig:Figure_5}c--d). The strongest loss occurred for the large negative perturbation. After Versatile Diffusion, PixCorr at $\alpha=-4$ fell to $0.17$ for valence and $0.26$ for memorability. Large positive perturbations also produced marked degradation, with PixCorr values of $0.35$ and $0.37$ at $\alpha=+4$, respectively. The corresponding CLIP similarities were higher and more stable (valence: $0.85$ and $0.87$; memorability: $0.86$ and $0.86$ at $\alpha=-4$ and $+4$, respectively), indicating that coarse semantic content could remain recognizable despite substantial changes in pixel-level structure.
The same boundary was already visible at the VDVAE stage, particularly for memorability. PixCorr declined to $0.47$ at $\alpha=-4$ and $0.59$ at $\alpha=+4$, while CLIP similarity declined to $0.79$ and $0.78$, respectively. For valence, VDVAE fidelity was somewhat better but still lower at the extremes (PixCorr $=0.68$ and $0.61$; CLIP similarity $=0.84$ and $0.81$ at $\alpha=-4$ and $+4$). Thus, increasing perturbation strength can produce larger automated cognitive-score changes while simultaneously moving the generated stimulus farther from its unperturbed reconstruction. The sharpest fidelity boundary occurred at $\alpha=-4$, with a second clear loss at $\alpha=+4$.
An exploratory condition-level analysis did not establish that fidelity statistically moderated the correspondence between assessor and human effects ($\beta=0.29$, bootstrap 95\% CI [$-0.22,0.78$]). Because this moderation analysis used condition-level fidelity summaries, it cannot resolve item-level moderation. The separate source-index-matched quality-filtering analysis in Appendix~\ref{app:qc_sensitivity} examines robustness to removing heavily altered stimuli, but is not a test of item-level moderation. We therefore treat attribute-score modulation, assessor-score preservation, and image fidelity as complementary outcomes rather than combining them into a single performance index. Together, Figures~\ref{fig:Figure_4} and~\ref{fig:Figure_5} suggest a practical validity region of the perturbation: moderate values provide a more favorable balance between targeted score change and visual preservation, whereas extreme values, especially $\alpha=-4$, increasingly enter a degraded and less interpretable stimulus regime.

Surrogate-predicted cortical maps provide a complementary, model-implied visualization of these perturbations (Appendix~\ref{subsec:neural_patterns}; Figure~\ref{fig:Figure_6}). They are presented separately from the behavioral and fidelity results because they are not independently measured neural responses or validated stimulation targets.

\section{Discussion}
\label{discussion}

\subsection{Methodological and Theoretical Context}

Our aim was to test whether an fMRI-derived AI surrogate could move beyond decoding to propose explicit, falsifiable representational changes and expose their predicted perceptual consequences to behavioral evaluation before physical stimulation. The results provide qualified support for this upstream role. The fMRI-to-latent mapping recovered coarse generative structure but not image-specific fine detail. At baseline, automated attribute structure was preserved after Versatile Diffusion, but human--assessor correspondence was numerically stronger for valence ($r=0.30$) than for perceived memorability ($r=0.10$), although the valence association remained uncertain ($p=0.058$). VDVAE perturbations then produced large and largely graded automated changes, but Versatile Diffusion attenuated or altered these responses. Human valence ratings followed the expected overall direction under the primary correction model, although the full-range effect was small and sensitive to alternative detrending and image-quality filtering; perceived memorability did not change systematically. Extreme perturbations also reduced reconstruction fidelity. The findings therefore support a bounded method for proposing and testing candidate representational changes, not a validated system for neural or clinical control.

\textbf{Representational alignment, prediction, and the Good Regulator principle.}
The approach builds on evidence that cognitive attributes can be locally separated within deep generative representations \citep{goetschalckx2019ganalyze,pmlr-v285-rothermel24a} and that artificial and biological visual systems share aspects of representational geometry \citep{conwell2023controlled,ozcelik2023natural,sucholutsky2023getting}. Such alignment does not imply common biological mechanisms. Rather, it provides a tractable correspondence through which candidate directions can be estimated in an artificial latent space and related to predicted neural patterns \citep{gaziv2023strong,bashivan2019neural}.

This distinction illustrates the relevance of the Good Regulator Theorem \citep{conant1970every}. Prediction and intervention are not equivalent: a model may predict an outcome without providing a useful account of how the represented system should be changed to alter that outcome. A model becomes relevant to regulation when it preserves the task-relevant relationships needed to propose interventions and anticipate their consequences. The present framework tests this intermediate requirement by asking whether an fMRI-derived surrogate can convert predictive structure into explicit, dose-dependent perturbation hypotheses whose perceptual consequences can be evaluated. It does not establish that the surrogate is itself a complete regulator, because it contains neither neural dynamics nor a causal mapping from stimulation inputs to neural responses.

The subject-specific mapping $R$ is central to the correspondence between measured fMRI patterns and the generative representation. The analyses in Appendix~\ref{role_of_R} show that it recovers coarse stimulus-related latent structure and defines a neurally decodable subspace. However, routing perturbations through that subspace did not improve the signed dose--response relative to direct latent steering and reduced image fidelity. These negative results are informative for the Good Regulator argument: preserving some predictive brain--latent structure is not sufficient to guarantee useful intervention directions. They clarify what the present surrogate captures, where it fails, and which additional causal and dynamical components will be required before it can support physical control.

\textbf{VDVAE and diffusion represent different surrogate-response regimes.}
The marked difference between the two reconstruction stages is theoretically
important. VDVAE showed the clearest dose-dependent score changes, whereas
Versatile Diffusion attenuated memorability modulation and largely compressed
the valence effect. At the same time, Versatile Diffusion preserved the relative
ordering of assessor scores reasonably well. Thus, rank-order preservation,
mean displacement, and visual fidelity capture different properties of the
pipeline: a stage can retain which images score higher while reducing the
magnitude of the intended representational perturbation.

This distinction is consistent with emerging concerns about highly realistic
brain-to-image reconstructions. Diffusion models can improve visual clarity and
semantic plausibility by imposing strong generative priors, including CLIP- or
text-derived information, but the added content need not be tightly determined
by the neural signal. This is a specific instance of a broader model--human
alignment problem: deep networks can rely on shortcut features that transfer
poorly, and optimization against learned visual models can produce images that
receive extreme model scores without remaining recognizable to humans
\citep{geirhos2020shortcut,nguyen2015deep}. Shirakawa and colleagues showed that text-guided diffusion
pipelines can generalize poorly across datasets and can generate realistic but
perceptually inaccurate images; they therefore cautioned that apparent realism
may partly reflect category inference and generative hallucination rather than
faithful reconstruction \citep{shirakawa2025spurious}. Recent reviews similarly
emphasize the need to separate the brain-to-latent translator, latent
representation, and generator when evaluating neural reconstruction
\citep{kamitani2025visual}. Our results provide a within-pipeline illustration
of this issue: diffusion produced visually coherent outputs, but changed the
attribute response inherited from VDVAE and substantially reduced pixel-level
continuity. This does not prove that VDVAE is intrinsically more ``neural'' than
diffusion. It does show that sharper final images are not sufficient evidence
of stronger neural fidelity and that steering effects must be evaluated at each
stage of the generative pipeline. This distinction is particularly important here because the perturbed Versatile Diffusion series used fixed CLIP embeddings from the corresponding original test image and caption; the refined outputs were therefore not end-to-end fMRI-only reconstructions.

\textbf{Local validity and the size of the human effect.}
The second limit concerns perturbation size. The theoretical formulation is a local, first-order approximation. Its validity therefore depends on perturbations remaining within a region in which the attribute direction, decoder, and generator behave approximately linearly. This
is the standard rationale for trust-region methods, which restrict optimization to a neighborhood in which the local approximation remains predictive
\citep{nocedal2006numerical}. Accordingly, the fidelity loss and stage-dependent response at $\alpha=\pm4$ identify a boundary of this regime rather than a monotonic increase in steering efficacy. One
plausible explanation for the smaller human valence effect is that the
behaviorally valid linear regime lies closer to $\alpha=0$ than the range used
here. Large automated shifts would then partly reflect movement into regions
where the assessor remains responsive but the generated stimulus is degraded
or its perceptual meaning changes. This explanation is consistent with the
present pattern, but it is not established by it: the experiment contained
only five widely spaced perturbation levels, and human estimates at individual
levels were uncertain. A decisive test requires denser sampling around zero,
for example $\alpha\in\{-2,-1,-0.5,0,0.5,1,2\}$, together with adaptive step-size
selection and prospective fidelity constraints.

The empirical direction $\boldsymbol{\theta}$ should accordingly be understood
as a population-level attribute axis, not an exact sample-wise gradient. It is
black-box compatible and can be estimated from scalar scores, including
non-differentiable or human assessments, but it does not guarantee locally
optimal movement for every image. The exact-gradient benchmark in Appendix~\ref{sec:exact_gradient_bound} directly illustrates this distinction: local gradients were nearly orthogonal to the population axis, and iterative per-image optimization produced larger assessor gains. These benchmark results do not replace the full-range behavioral analysis; they clarify why a shared population axis is not a locally optimal direction for every image. Exact gradients remain locally valid and do not eliminate generator-induced nonlinearities.
Future implementations should therefore combine local directions with
trust-region constraints, nonlinear optimization, or iterative feedback rather
than extrapolating a single linear update to large perturbations.

\textbf{Valence and perceived memorability.}
Across the tested perturbation range, independent human evaluation provides the critical test. The behavioral results differed between attributes, alongside the numerically different but uncertain baseline correspondence in Section~\ref{sec:baseline-agreement}. The human valence result was
directionally consistent across most participants, but its magnitude was much
smaller than the automated effect and strict monotonicity was uncommon. It
should therefore be interpreted as preliminary evidence that the perturbation
affected ratings of image valence under the stated correction model, not as a final validation of a reliable human
dose--response function.

For memorability, the experimental limitation is more fundamental. Participants
judged how memorable an image appeared immediately after presentation; they did
not complete a subsequent-memory test. Prior work likewise distinguishes
subjective memorability judgments from actual recognition performance and shows
that their correspondence depends on the cues used to make those judgments
\citep{shoval2023objects}. Consistent with this distinction, the judgment
in our study was only weakly related to
the automated memorability score at baseline ($r=0.10$;
Table~\ref{tab:baseline_agreement}). The absence of a human perturbation effect therefore does not show
that memory itself cannot be modulated, nor does it validate the automated
memorability changes. It shows that the perturbations did not systematically
alter perceived memorability under the present task. Future work should use an
incidental-encoding design followed by delayed recognition or recall and should
evaluate trial-level subsequent-memory outcomes alongside perceived
memorability.

Additional limitations affect both attributes. The behavioral conditions were
presented in a fixed block order, confounding condition with
time and possible fatigue. Leave-one-participant-out correction and the sensitivity
analyses in Appendix~\ref{app:behavioral_sensitivity} reduce this concern but cannot remove it completely. The automated
analyses were based on four fMRI participants, and the cortical maps were
surrogate-derived predictions rather than independently measured activation
changes. Finally, the fidelity-moderation analysis used condition-level summaries. The source-index-matched filtering analysis in Appendix~\ref{app:qc_sensitivity} showed that the full-range valence estimate depended on the retained stimulus set, but the missing fMRI-subject provenance of the rated reconstructions prevents an exact rendered-image match. Filtering also reduces precision and can remove an entire perturbation level. These checks therefore do not establish that image degradation caused the human--assessor divergence.

\textbf{From a static approximation to dynamical control.}
These behavioral results define the endpoint of the present evidence; moving from perceptual evaluation to neural intervention requires a dynamical model. Control theory provides the general framework within which the present target-identification problem is situated, but the implementation studied here is intentionally static. Stimulus-evoked BOLD patterns are mapped to latent representations, and a single representational perturbation is selected under a norm constraint. This corresponds to choosing a candidate local direction at one operating point, not to controlling an evolving biological state.

Cognitive and affective states instead evolve through recurrent and state-dependent dynamics. A fuller formulation would introduce

$$
\mathbf{x}_{t+1}=f(\mathbf{x}_t,\mathbf{u}_t),\qquad
\mathbf{y}_t=g(\mathbf{x}_t),
$$

where $\mathbf{u}_t$ denotes a feasible stimulation input, $\mathbf{x}_t$ the evolving neural state, and $\mathbf{y}_t$ the cognitive-affective outcome. Such a model would permit optimization over trajectories rather than isolated endpoints \citep{kheirkhah2026re} and could represent temporal integration, hysteresis, multistability, feedback, and compensatory responses \citep{durstewitz2023reconstructing,luo2025frequency}. The present study addresses the preceding question of which representational change should be sought and whether its predicted perceptual consequence survives behavioral evaluation. It should therefore be viewed as a simplified upstream component of a broader control-theoretic program, not as a demonstration of dynamical or closed-loop neural control.

\subsection{Translational and Ethical Implications}

The main translational contribution is a procedure for proposing and testing
candidate representational changes before attempting stimulation. It
shifts attention from anatomical targeting alone toward the functional pattern
associated with a desired cognitive change
\citep{kheirkhah2026re,bosl2025dynamical}. However, no neural perturbation was
delivered in this study, and the predicted cortical maps are neither empirical
activation effects nor validated stimulation targets. Whether the distributed
patterns can be approximated by non-invasive hardware, whether compensatory
network dynamics would oppose them, and whether they generalize beyond visual
cortex remain open questions.

These uncertainties define a substantial hardware-alignment gap. The inferred
directions are high-dimensional and spatially distributed, whereas current
non-invasive stimulation has limited spatial specificity. Network-control work
also suggests that transitions induced from a small number of control sites can
be energetically demanding \citep{suweis2019brain}. High-density, multi-site,
and closed-loop systems may eventually approximate distributed control signals,
but this requires direct prospective testing rather than inference from the
present reconstructions \citep{scangos2021closed,guidotti2025neuromodulation}.

The same caution applies to clinical interpretation. A directional change in
valence ratings does not demonstrate correction of pathological affective
states, and the current effect size should not be equated with therapeutic
benefit. The framework is instead a means of making candidate representational targets explicit, dose-dependent, falsifiable, and testable against both behavioral and
neural outcomes. Clinical translation will require randomized perturbation or
stimulation studies, individualized models, trajectory-level endpoints, and
evidence that changes persist and improve clinically meaningful functioning.

Finally, representational steering raises concerns about autonomy, consent,
privacy, and misuse. Any progression from in silico design to intervention
should require transparent objectives, auditable models, participant control,
and regulation proportionate to risk \citep{yuste2017four,ienca2017towards}.
These safeguards are especially important because generative outputs can appear
more precise than the neural evidence supporting them. Responsible development
therefore depends as much on communicating failure modes and uncertainty as on
improving reconstruction quality.

\section{Conclusion}

This study presents a control-theory-motivated framework that implements a simplified, static inverse-design step through an fMRI-derived AI surrogate. The broader control problem is to determine how feasible interventions can alter evolving neural states to produce desired cognitive-affective outcomes; the present work addresses the upstream question of which candidate representational changes should be considered and whether their predicted consequences survive behavioral evaluation. Perturbations produced large, graded changes in automated scores at the VDVAE stage, but these effects were altered by Versatile Diffusion. Human valence ratings followed the expected overall direction under the primary correction model, although the effect was smaller than the automated response, was not consistently monotonic, and was sensitive to alternative correction specifications and image-quality filtering. No corresponding effect was detected for perceived memorability, which showed weak baseline correspondence with the automated memorability assessor. Together with declining visual fidelity at extreme perturbation strengths, these findings define the validity limits of the present static approximation. The main contribution is therefore a falsifiable framework for testing candidate representational targets within a broader control-theoretic approach, rather than a demonstration of dynamical, neural, or clinical control. Future work must connect these targets to causal stimulation models, neural dynamics, feedback, and prospective behavioral validation.

\backmatter

\section{Declarations}

\subsection{Ethical Approval}  
All participants provided written informed consent prior to their participation in the experiments. The study protocols were reviewed and approved by the Ethics Committee of the University of Marburg.  

\subsection{Funding Acknowledgment}  
This work was supported in part by the German Research Foundation (DFG) through the Collaborative Research Center SFB/TRR 393 (project grant no.\ 521379614), by a research grant from the von Behring-Röntgen Stiftung (grant no.\ 70\_00038), by a research grant from the University Hospital of Gießen and Marburg (UKGM, grant no.\ 1/2024 MR), and by the Excellence Cluster EXC3066 “The Adaptive Mind”. This paper is further partially supported by ERA-NET NEURON JTC 2024 (BRAWO project, grant no. ER-2024-23684536). The funders had no role in study design, data collection, management, analysis, interpretation, report writing, or the decision to submit the manuscript for publication. The authors retained full responsibility for the decision to submit.  

\subsection{Data and Code Availability}
\label{sec:data-code-availability}
All fMRI data used in this paper are publicly available at \url{https://naturalscenesdataset.org/}. All code used to generate the simulations and results will be made publicly available upon publication at \url{https://www.neuralsurrogate.com/}.  

\subsection{Conflict of Interest}  
H.J. is an Associate Editor at npj Mental Health Research and was not involved in the journal's review of, or decisions related to, this manuscript.

\subsection{Author Contributions}  
Conceptualization, Methodology, and Validation: MR, MS, SD, SGH, HJ.  
Statistical Analysis: MR, HJ.  
Data Acquisition: SJF, BS, JCGA.  
Writing -- Original Draft: MR, MS, HJ.  
Writing -- Review and Editing: all authors.


\newpage

\bibliography{sn-bibliography}

\newpage
\begin{appendices}
\appendix
\counterwithin{equation}{section}
\counterwithin{figure}{section}
\counterwithin{table}{section}
\renewcommand{\theequation}{\thesection\arabic{equation}}
\renewcommand{\thefigure}{\thesection\arabic{figure}}
\renewcommand{\thetable}{\thesection\arabic{table}}

\section{Reconstruction Implementation and Model Settings}
\label{app:reconstruction_implementation}
The reconstruction pipeline comprised three stages: subject-specific prediction of VDVAE and CLIP representations from fMRI, decoding of the predicted or perturbed VDVAE latents into coarse images, and refinement with Versatile Diffusion. Only the subject-specific ridge regressions were fitted in the present study. VDVAE, CLIP, Versatile Diffusion, EmoNet, and MemNet were pretrained and used with frozen parameters.

\subsection{fMRI preprocessing and ridge mappings}
\label{ridge reconstruction}
For each NSD subject $s$, the training and held-out fMRI matrices were
\begin{equation*}
X_{\mathrm{train}}^{(s)} \in \mathbb{R}^{8859 \times d_s}, \qquad X_{\mathrm{test}}^{(s)} \in \mathbb{R}^{982 \times d_s},
\end{equation*}
where $d_s$ denotes the number of voxels in the subject-specific NSD-general region of interest. Before regression, beta values were divided by 300 and standardized voxelwise using the training-set mean and standard deviation. The same parameters were applied to the held-out data. All mappings were fitted separately for each subject and included an intercept.

Let $Y_k \in \mathbb{R}^{N \times q_k}$ denote the target representation for reconstruction stream $k$. The multi-output ridge model was written consistently as
\begin{equation}
(\widehat{W}_k^{(s)},\widehat{\mathbf b}_k^{(s)}) =
\arg\min_{W,\mathbf b}
\left\|
Y_{k,\mathrm{train}} -
X_{\mathrm{train}}^{(s)}W -
\mathbf{1}\mathbf b^\top
\right\|_F^2
+
\lambda_k \|W\|_F^2,
\label{eq:appendix_ridge}
\end{equation}
with $W_k^{(s)} \in \mathbb{R}^{d_s \times q_k}$. Held-out representations were predicted as
\begin{equation}
\widehat{Y}_{k,\mathrm{test}}^{(s)} =
X_{\mathrm{test}}^{(s)}\widehat{W}_k^{(s)}
+
\mathbf{1}\widehat{\mathbf b}_k^{(s)\top}.
\label{eq:appendix_prediction}
\end{equation}
For the theoretical notation used in the main text, the fMRI-to-VDVAE mapping is $R^{(s)} = \widehat{W}_{\mathrm{VDVAE}}^{(s)\top}$, so that $R^{(s)} \in \mathbb{R}^{91{,}168 \times d_s}$.

The regression settings are summarized in Table~\ref{tab:regression_hyperparameters}.
\begin{table}[htbp]
\centering
\caption{\textbf{Parameters of the subject-specific fMRI-to-feature regressions.}
All models included an intercept and used training-set fMRI normalization.}
\label{tab:regression_hyperparameters}
\begin{tabular}{lccc}
\toprule
Regression target & Output per image & Ridge penalty $\lambda$ & Maximum iterations \\
\midrule
VDVAE latents
& 91{,}168
& 50{,}000
& 10{,}000 \\
CLIP text embeddings
& $77 \times 768$
& 100{,}000
& 50{,}000 \\
CLIP vision embeddings
& $257 \times 768$
& 60{,}000
& 50{,}000 \\
\bottomrule
\end{tabular}
\end{table}

For example, subject 1 contributed 15{,}724 NSD-general voxels, giving a VDVAE regression matrix of $15{,}724 \times 91{,}168$ and a bias vector of length 91{,}168. The voxel dimension differed across subjects.

\subsection{VDVAE representation and decoding}
\label{vae recon}
Stimulus images were encoded with the frozen ImageNet-64 VDVAE checkpoint \texttt{imagenet64-iter-1600000} \citep{child2020very}. The model contains 31 hierarchical latent layers, which were concatenated into a 91{,}168-dimensional vector. These image-derived vectors constituted the targets for the fMRI-to-VDVAE ridge regression.

Before image decoding, the raw held-out VDVAE predictions were rescaled componentwise to the training-latent distribution:
\begin{equation}
\widehat{Z}_{\mathrm{decode}}^{(s)} =
\boldsymbol{\mu}_{\mathrm{train}}
+
\boldsymbol{\sigma}_{\mathrm{train}}
\odot
\frac{
\widehat{Z}_{\mathrm{raw}}^{(s)} -
\boldsymbol{\mu}_{\mathrm{pred}}^{(s)}
}{
\boldsymbol{\sigma}_{\mathrm{pred}}^{(s)}
},
\label{eq:appendix_latent_rescaling}
\end{equation}
where $\boldsymbol{\mu}_{\mathrm{pred}}^{(s)}$ and $\boldsymbol{\sigma}_{\mathrm{pred}}^{(s)}$ were calculated across the held-out prediction batch, and $\boldsymbol{\mu}_{\mathrm{train}}$ and $\boldsymbol{\sigma}_{\mathrm{train}}$ were calculated from the training-image latents. This operation used no held-out target latents, but it was transductive because the rescaled representation of one test image depended on the other predictions in the held-out batch. Consequently, all latent-prediction fidelity analyses reported in Appendix~\ref{role_of_R} were calculated from the raw ridge predictions before this rescaling.

The cognitive-attribute series was constructed from the rescaled fMRI-decoded latent:
\begin{equation}
\widehat{\mathbf z}_{\alpha}^{(s)} =
\widehat{\mathbf z}_{0}^{(s)}
+
\alpha
\frac{\boldsymbol{\theta}}
{\|\boldsymbol{\theta}\|_2},
\qquad
\alpha \in \{-4, -2, 0, +2, +4\},
\label{eq:appendix_alpha_series}
\end{equation}
where $\boldsymbol{\theta}$ was the difference between the mean VDVAE latent vectors of the highest- and lowest-scoring training-image quartiles. Perturbations were applied directly in VDVAE latent space; the projected fMRI-space direction $R^\top\boldsymbol{\theta}$ was used only for surrogate-based visualization and was not applied to measured brain activity.

The frozen VDVAE decoder produced a native $64 \times 64$-pixel RGB image:
\begin{equation}
\widehat{I}_{\alpha,\mathrm{VDVAE}}^{(s)} =
\mathcal{D}_{\mathrm{VDVAE}}
\left(
\widehat{\mathbf z}_{\alpha}^{(s)}
\right)
\in \mathbb{R}^{64 \times 64 \times 3}.
\label{eq:appendix_vdvae_decode}
\end{equation}
These outputs were resized to $512 \times 512$ pixels before assessor evaluation and diffusion refinement. 

\subsection{CLIP conditioning representations}
\label{clip reconstruction 1}
CLIP features were extracted using the CLIP encoders distributed with Versatile Diffusion \citep{radford2021learning,xu2023versatile}. The text stream produced a $77 \times 768$ token representation. Each NSD image had multiple human-written COCO captions; captions were encoded separately and their representations were averaged to obtain one text-conditioning representation per image. No captions were generated by the reconstruction model.

The vision stream produced a $257 \times 768$ patch-token representation per image. For ridge regression, the text and vision representations were separately vectorized, predicted using Equation~\ref{eq:appendix_ridge}, and reshaped to their original token dimensions before diffusion. Because the two streams contain different numbers of tokens, they were not added together as latent vectors.

The source of the CLIP conditioning depended on the reconstruction analysis:
\begin{enumerate}
\item \textbf{End-to-end baseline brain reconstruction.}
CLIP text and vision representations were predicted from held-out fMRI activity using the subject-specific ridge mappings. No original test-image or ground-truth caption representation was provided to the diffusion model.
\item \textbf{Controlled \(\alpha\)-series reconstruction.}  
The CLIP text and vision representations were extracted from the corresponding original test image and its COCO captions and then held fixed across all five values of \(\alpha\), including the \(\alpha=0\) reference. Thus, the VDVAE scaffold remained fMRI-derived, but the semantic diffusion conditioning was source-matched rather than predicted from fMRI.
\end{enumerate}
The second design reduced semantic variation unrelated to the VDVAE perturbation and enabled controlled comparison across $\alpha$. However, it also means that the Versatile Diffusion $\alpha$-series should not be described as an entirely end-to-end fMRI reconstruction.

\subsection{Versatile Diffusion refinement}
\label{clip reconstruction 3}
The coarse VDVAE image was refined using the frozen \texttt{vd-four-flow-v1-0-fp16-deprecated} Versatile Diffusion checkpoint \citep{xu2023versatile}. Following the low-level/high-level reconstruction strategy of \citet{ozcelik2023natural}, the resized VDVAE output supplied the initial image scaffold, while the CLIP vision and text streams supplied separate semantic-conditioning signals.

For each perturbation level, the resized VDVAE image was first encoded into the AutoKL latent space:
\begin{equation}
\mathbf h_{\alpha,0}^{(s)} =
\mathcal{E}_{\mathrm{AutoKL}}
\left(
\operatorname{Resize}_{512}
\left[
\widehat{I}_{\alpha,\mathrm{VDVAE}}^{(s)}
\right]
\right).
\label{eq:appendix_autokl_encode}
\end{equation}
Noise was introduced according to the image-to-image diffusion strength, after which DDIM sampling was performed using the two conditioning branches:
\begin{equation}
\mathbf h_{\alpha,\mathrm{final}}^{(s)} =
\operatorname{DDIM}
\left(
\mathbf h_{\alpha,t}^{(s)};
\mathbf c_{\mathrm{vision}},
\mathbf c_{\mathrm{text}},
w_{\mathrm{vision}}=0.8,
w_{\mathrm{text}}=0.2
\right).
\label{eq:appendix_vd_sampling}
\end{equation}
Here, $w_{\mathrm{vision}}$ and $w_{\mathrm{text}}$ describe the model's separate conditioning mixture; they do not imply direct addition of the differently shaped CLIP representations. Sampling used 50 DDIM steps, $\eta=0$, classifier-free guidance scale 20, diffusion strength 0.5, and a text--vision mixing value of 0.2, corresponding to the 0.8/0.2 vision/text ratio.

The final image was obtained with the frozen AutoKL decoder:
\begin{equation}
\widehat{I}_{\alpha,\mathrm{VD}}^{(s)} =
\mathcal{D}_{\mathrm{AutoKL}}
\left(
\mathbf h_{\alpha,\mathrm{final}}^{(s)}
\right)
\in \mathbb{R}^{512 \times 512 \times 3}.
\label{eq:appendix_vd_decode}
\end{equation}
All final reconstructions were stored as $512 \times 512$-pixel PNG images. VDVAE and Versatile Diffusion were not retrained or fine-tuned on NSD. EmoNet and MemNet were likewise used only as frozen assessors. Therefore, epochs and learning rates apply to none of these pretrained networks; the fitted parameters in this study were limited to the subject-specific ridge mappings described above.

\subsection{Computation of Surrogate-Derived Cortical Maps}
\label{app:cortical_map_computation}
Figure~\ref{fig:Figure_6} visualizes how the latent attribute directions are projected into the voxel space represented by the fMRI-derived surrogate. The maps were computed separately for valence and memorability and for the extreme perturbation strengths $\alpha=-4$ and $\alpha=+4$. No additional model was fitted for this visualization.

For subject $s$, the fitted ridge decoder
\begin{equation*}
R^{(s)} \in \mathbb{R}^{n \times d_s}
\end{equation*}
maps a voxelwise fMRI pattern $\mathbf{x}^{(s)} \in \mathbb{R}^{d_s}$ to the VDVAE latent representation $\mathbf{z} \in \mathbb{R}^{n}$, where $n = 91{,}168$. For attribute $k$, with empirical latent direction $\boldsymbol{\theta}_k$, the corresponding voxel-space sensitivity direction was obtained using the adjoint of the decoder:
\begin{equation}
\mathbf{v}_k^{(s)} =
R^{(s)\top}
\frac{\boldsymbol{\theta}_k}
{\|\boldsymbol{\theta}_k\|_2}.
\label{eq:appendix_voxel_direction}
\end{equation}
The candidate change associated with perturbation strength $\alpha$ was then
\begin{equation}
\Delta\mathbf{x}_{k,\alpha}^{(s)} =
\alpha
\frac{\mathbf{v}_k^{(s)}}
{\|\mathbf{v}_k^{(s)}\|_2},
\qquad
\alpha \in \{-4, +4\}.
\label{eq:appendix_voxel_perturbation}
\end{equation}
This operation is a linear back-projection of the latent attribute direction through the fitted statistical decoder. It is not an inversion of the neural system and does not estimate a physically realizable stimulation input.

To express the displayed values relative to the unperturbed condition, the model-implied pattern for each perturbation was compared with its corresponding $\alpha=0$ pattern. For voxel $j$, the signed relative contrast was calculated as
\begin{equation}
C_{k,\alpha,j}^{(s)} =
\frac{
\widehat{\beta}_{k,\alpha,j}^{(s)} -
\widehat{\beta}_{k,0,j}^{(s)}
}{
\left|
\widehat{\beta}_{k,\alpha,j}^{(s)}
\right|
+
\left|
\widehat{\beta}_{k,0,j}^{(s)}
\right|
+
\varepsilon
},
\label{eq:appendix_relative_contrast}
\end{equation}
where $\varepsilon$ is a small numerical constant preventing division by zero. This normalization was used only to place voxelwise changes on a comparable descriptive scale; it does not produce a statistical contrast or $t$-map.

The signed contrast was separated into positive and negative display components:
\begin{equation}
C_{k,\alpha,j}^{(s)+} =
\max\left(C_{k,\alpha,j}^{(s)},0\right),
\qquad
C_{k,\alpha,j}^{(s)-} =
\max\left(-C_{k,\alpha,j}^{(s)},0\right).
\label{eq:appendix_signed_components}
\end{equation}
The positive component was displayed in yellow and the negative component in cyan. Thus, negative contrasts were not discarded from the analysis: values of the opposite sign were set to zero only within each separate color layer.

Voxelwise values were inserted into the native functional volume using the subject-specific NSD-general voxel mask and saved in NIfTI format. Because the decoder was fitted only to voxels within this mask, no values were extrapolated to unmodelled brain regions. Each volume was aligned with the corresponding anatomical T1-weighted image and projected onto the cortical surface using FreeSurfer \citep{fischl2000measuring,fischl1999b}. The surface rendering was used exclusively for anatomical visualization.

No general linear model, group-level statistical contrast, cluster correction, or vertexwise inferential test was applied. The displayed maps are therefore descriptive sensitivity maps derived from the fitted surrogate. They should not be interpreted as measured activation during perturbation, evidence of regional target engagement, or validated stimulation targets.

\FloatBarrier
\section{The Role of the fMRI-to-Latent Mapping \(R\)}
\label{role_of_R}

The subject-specific mapping \(R\) projects measured fMRI patterns into the VDVAE latent space. We examined two distinct questions: whether \(R\) recovers stimulus-related latent information from fMRI and whether routing perturbations through the subspace defined by \(R\) improves cognitive-attribute steering. These questions should not be conflated: successful fMRI decoding does not necessarily imply improved steering or causal controllability.

\subsection{Held-out latent decodability}

Latent reconstruction fidelity was evaluated on the fixed set of 982 held-out images for each NSD subject. Predictions were compared with the VDVAE latents obtained by encoding the corresponding stimulus images. All metrics were calculated from the raw ridge predictions before the batch-dependent rescaling used for image reconstruction.

Because VDVAE is hierarchical, decodability was evaluated separately by latent resolution (Table~\ref{tab:layer_resolved_fidelity}). Held-out \(R^2\) was defined relative to a predictor that assigns every test image the training-set mean latent vector. Thus, \(R^2=0\) represents the training-mean baseline and negative values indicate performance below that baseline.

\begin{table}[htbp]
\centering
\caption{\textbf{Layer-resolved held-out \(R^2\) of the fMRI-to-VDVAE mapping.}
Values are reported relative to a training-mean latent predictor. Coordinates denote the number of latent variables within each spatial-resolution group.}
\label{tab:layer_resolved_fidelity}
\begin{tabular}{lrrrrr}
\toprule
Resolution & Coordinates & S1 & S2 & S5 & S7 \\
\midrule
$1\times1$   & 32       & 0.252  & 0.230  & 0.243  & 0.200 \\
$4\times4$   & 1{,}024  & 0.187  & 0.149  & 0.117  & 0.117 \\
$8\times8$   & 8{,}192  & $-0.011$ & $-0.012$ & $-0.015$ & $-0.013$ \\
$16\times16$ & 65{,}536 & $-0.022$ & $-0.021$ & $-0.019$ & $-0.019$ \\
$32\times32$ & 16{,}384 & $-0.023$ & $-0.022$ & $-0.019$ & $-0.019$ \\
\bottomrule
\end{tabular}
\end{table}

The learned mapping recovered the coarse \(1\times1\) and \(4\times4\) latent components above baseline in every subject. Two-way identification reached \(0.79\)--\(0.88\) for layer 4 and \(0.80\)--\(0.86\) for layer 6, compared with chance performance of \(0.5\). In contrast, the finer-scale latent components were not reliably predicted. Because these components comprise most of the \(91{,}168\)-dimensional representation, coordinate-weighted \(R^2\) was negative (\(-0.020\) to \(-0.017\)), whereas weighting the 31 hierarchical layers equally yielded positive values of \(0.010\)--\(0.020\). The layer-resolved results therefore provide the more informative description: fMRI supported recovery of coarse latent structure but not image-specific fine detail.

This signal depended on the correspondence between fMRI activity and the viewed image. Permuting the held-out fMRI--image pairing or shuffling the training-image--latent pairing reduced coordinate correlations to approximately zero and two-way identification to \(0.47\)--\(0.53\). Ridge models fitted to Gaussian random features with matched rank retained some low-rank predictive structure but performed more weakly than the learned fMRI mapping: \(4\times4\)-layer \(R^2\) was \(0.079\)--\(0.126\), compared with \(0.117\)--\(0.187\) for the learned mapping, and layer-4 identification was \(0.75\)--\(0.83\), compared with \(0.79\)--\(0.88\). A rank-\(1{,}000\) PCA-plus-ridge model performed similarly to the full mapping, indicating that the decodable fMRI signal was predominantly low-rank.

\subsection{Does \(R\) improve latent steering?}

The automated perturbation effects reported in the main analysis were generated by applying \(\boldsymbol{\theta}\) directly to the rescaled fMRI-decoded latent representations. We first tested whether the resulting score changes depended specifically on the learned mapping used to generate the unperturbed latent. One hundred fixed held-out images per subject were reconstructed using base latents obtained from the learned mapping, PCA-plus-ridge, or five random-projection-plus-ridge controls. Decoder sampling noise was held identical across conditions.

Memorability increased locally with \(\alpha\) for all three base-latent mappings. However, the paired differences between learned and control mappings were small, with a maximum absolute difference of approximately \(0.005\) assessor-score units, and no advantage of the learned mapping replicated across subjects. Valence was non-monotonic for the raw latent predictions under all mapping conditions. This comparison indicates that the automated steering profile was primarily determined by the attribute direction, the latent-space operating point, and the nonlinear decoder---not by the detailed structure of the mapping that generated the base latent.

We next compared direct latent steering with neural-routed steering (Table~\ref{tab:direct_neural_routing}). Direct steering used

$$
\Delta\mathbf z_{\mathrm{direct}}
\propto
\boldsymbol{\theta},
$$

whereas neural-routed steering used

$$
\Delta\mathbf z_{\mathrm{neural}}
\propto
RR^\top\boldsymbol{\theta}.
$$

The second operation maps the attribute direction through fMRI space and back into latent space, reweighting it according to the learned decoder; $RR^\top$ is not generally an orthogonal projector. Direct, neural-routed, row-permuted-\(R\), and Gaussian-random routes were matched for realized latent perturbation norm. The same base images, attribute directions, and decoder-noise realizations were used in every condition.

\begin{table}[htbp]
\centering
\caption{\textbf{Comparison of direct and neural-routed latent perturbations.}
Ranges summarize the four NSD subjects and the two cognitive attributes where applicable. PixCorr was computed relative to the corresponding unperturbed reconstruction.}
\label{tab:direct_neural_routing}
\begin{tabular}{lccc}
\toprule
Measure
& Direct \(\boldsymbol{\theta}\)
& Neural \(RR^\top\boldsymbol{\theta}\)
& Permuted/random routing \\
\midrule
Cosine with \(\boldsymbol{\theta}\)
& \(1.00\)
& \(0.30\)--\(0.37\)
& -- \\
Energy in \(1\times1\) and \(4\times4\) blocks
& \(3\%\)--\(9\%\)
& \(23\%\)--\(45\%\)
& \({<}\,3\%\) \\
PixCorr with \(\alpha=0\)
& \(0.60\)--\(0.86\)
& \(0.13\)--\(0.24\)
& \(0.92\)--\(0.98\) \\
Memorability: images with intended \(0\rightarrow+2\) change
& \(65\%\)--\(92\%\)
& \(0\%\)--\(16\%\)
& Not consistently observed \\
Subjects with monotonic memorability
& \(4/4\)
& \(0/4\)
& \(0/4\) \\
\bottomrule
\end{tabular}
\end{table}

The learned neural route concentrated \(23\%\)--\(45\%\) of the perturbation energy in the coarse \(1\times1\) and \(4\times4\) latent blocks, compared with \(3\%\)--\(9\%\) for direct steering and less than \(3\%\) for permuted and random routes. Thus, \(R\) identified a specific neurally decodable subspace. The neural route also produced larger assessor-score changes than the permuted-\(R\) controls in all eight subject-by-target comparisons.

Larger score changes did not, however, correspond to better directional control. Neural routing produced substantially greater image alteration, with PixCorr values of \(0.13\)--\(0.24\), compared with \(0.60\)--\(0.86\) for direct steering. For memorability, the positive neural-routed arm reversed in every subject, whereas direct steering retained monotonic condition means in all four subjects. The larger neural-routed score changes therefore reflected, at least partly, greater movement away from the original image rather than more accurately directed modulation. The image-quality sensitivity analysis in Appendix~\ref{app:qc_sensitivity} additionally shows that the net score advantage of neural routing over direct steering was threshold-dependent, whereas the intermediate-level direct-route comparison was unchanged under the frozen rule.

These analyses define the contribution of \(R\) more precisely. First, \(R\) recovers coarse, stimulus-related latent structure from fMRI above shuffled and random controls. Second, \(RR^\top\boldsymbol{\theta}\) concentrates perturbations in this decodable latent subspace. Third, this routing does not improve the signed dose--response relative to direct latent editing and can substantially reduce image fidelity. Accordingly, the primary perturbation results should be interpreted as direct latent steering applied at fMRI-decoded operating points, not as evidence that neural routing through \(R\) improves steering or provides a causal control direction.

\section{Validation of the Empirical Steering Direction}
\label{app:theta_validation}

The empirical steering direction for attribute $k$ was defined from the
training images as the difference between the mean VDVAE latent
representations of the highest- and lowest-scoring quartiles:
\begin{equation}
\boldsymbol{\theta}_k =
\overline{\mathbf z}_{k,\mathrm{high}} -
\overline{\mathbf z}_{k,\mathrm{low}}.
\label{eq:appendix_theta_validation}
\end{equation}
This direction was estimated separately for EmoNet valence and MemNet
memorability. To determine whether it generalized beyond the images used
for its construction, we projected the held-out test-image latents onto
the normalized direction:
\begin{equation}
q_{i,k} =
\mathbf z_i^\top
\frac{\boldsymbol{\theta}_k}
{\|\boldsymbol{\theta}_k\|_2}.
\label{eq:appendix_theta_projection}
\end{equation}
The held-out images were then categorized for visualization as low
scoring (bottom 15\%), middle scoring (central 70\%), or high scoring
(top 15\%) according to the corresponding frozen assessor. These
held-out categories were not used to estimate $\boldsymbol{\theta}_k$.

\begin{figure*}[t]
\centering
\includegraphics[width=\textwidth]{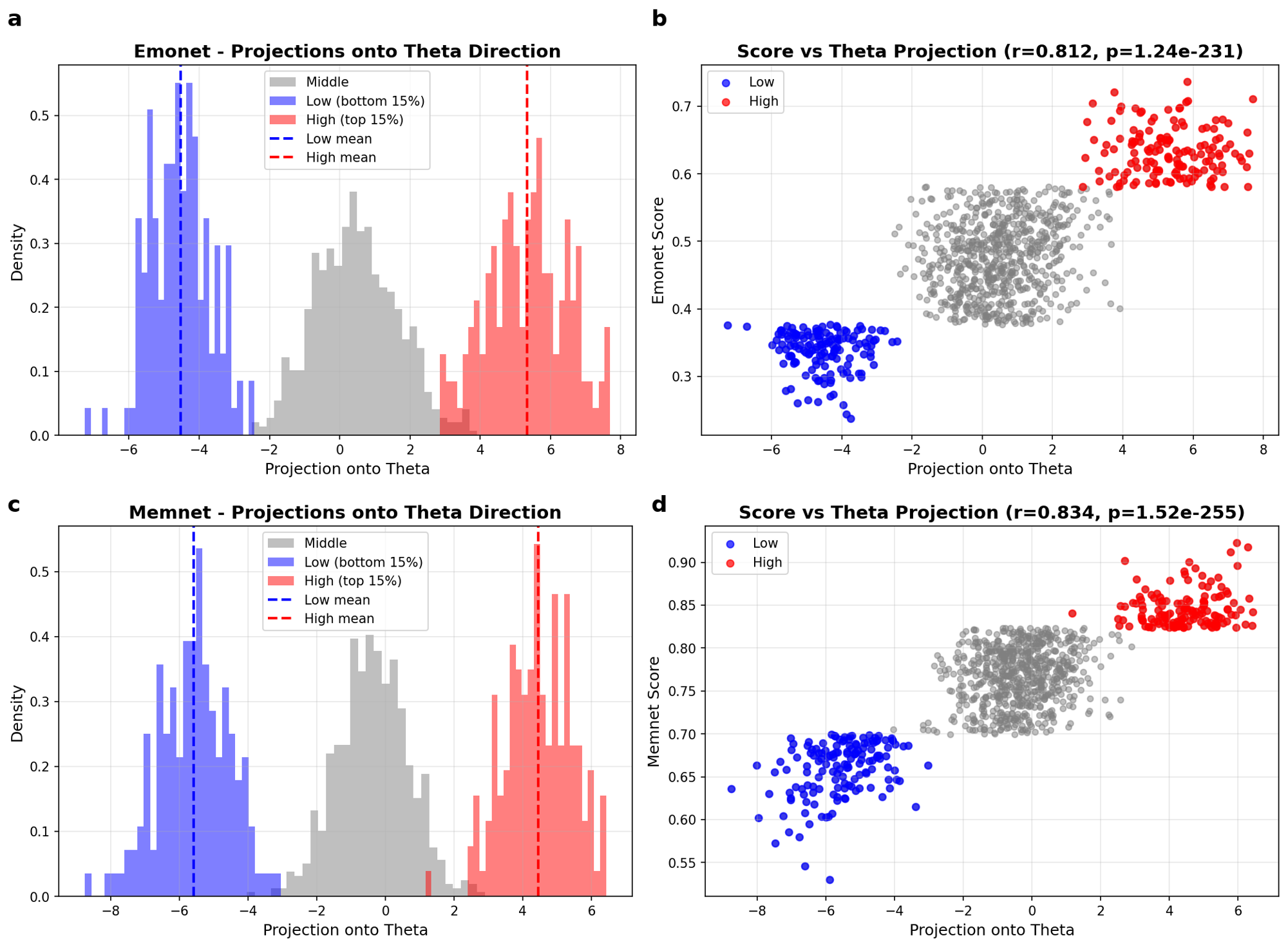}
\caption{\textbf{Held-out validation of the empirical steering directions.}
\textbf{(a)} Distribution of held-out VDVAE latent projections onto the
EmoNet valence direction for images in the bottom 15\%, central 70\%,
and top 15\% of held-out EmoNet scores. Dashed lines mark the mean
projections of the low- and high-scoring groups.
\textbf{(b)} Held-out EmoNet scores versus projection onto the valence
direction ($r=0.812$).
\textbf{(c)} Corresponding projection distributions for the MemNet
memorability direction.
\textbf{(d)} Held-out MemNet scores versus projection onto the
memorability direction ($r=0.834$).
The panels show fMRI subject 1; each direction was estimated from training-image latents only. Pearson
correlations quantify held-out geometric correspondence with the frozen
assessor; they do not constitute human or neural validation.}
\label{fig:appendix_theta_validation}
\end{figure*}

The projections separated low- and high-scoring held-out images for both
attributes (Figure~\ref{fig:appendix_theta_validation}a,c). Projection
values tracked the corresponding assessor scores for valence
($r=0.812$; Figure~\ref{fig:appendix_theta_validation}b) and
memorability ($r=0.834$;
Figure~\ref{fig:appendix_theta_validation}d). Thus, the
centroid-difference direction captured a reproducible global attribute
axis in VDVAE latent space rather than only separating the training
images from which it was estimated.

This analysis provides geometric validation of the empirical direction,
but its interpretation is limited. It does not establish that
$\boldsymbol{\theta}_k$ is the locally optimal gradient for every image, that
it represents a causal neural direction, or that movement along it must
produce a corresponding change in human judgments. Instead, it shows
that a single population-level direction preserves the ordering of
assessor-defined attributes in held-out latent representations.

\FloatBarrier
\subsection{Exact-gradient benchmark for the population steering direction}
\label{sec:exact_gradient_bound}

To calibrate the population direction $\boldsymbol{\theta}$ against image-specific optimization alternatives, we computed exact automatic-differentiation gradients of a deterministic surrogate objective with respect to the full 91{,}168-dimensional latent vector. The surrogate applied the assessor to the mixture-mean VDVAE decode, with the reparameterized prior draws for the decoder blocks beyond the 31 supplied latents fixed by a per-image seed. These are exact gradients of the deterministic surrogate, not of the stochastic sampling pipeline. The analysis was prespecified and version-controlled before computation.

For 20 predeclared held-out images per fMRI subject, we compared four steering strategies at matched realized latent norms of $1\times$ and $2\times$ $\|\boldsymbol{\theta}\|$ in this benchmark: a single exact-gradient step; five-step projected gradient ascent within the same norm ball, which serves as an empirical per-image optimization benchmark rather than a proven upper bound on the nonlinear objective; the frozen $\boldsymbol{\theta}$; and three random unit directions. Every condition was evaluated with the standard stochastic decoding and scoring pipeline under matched decoder noise, so that directions differed while evaluation was identical. No gradient or decoding failures occurred in the 160 image--target computations.

Three results follow. First, the assessor objectives are strongly steerable when optimized per image: projected gradient ascent raised scores in 95\%--100\% of images in every subject (EmoNet $+0.15$ to $+0.22$ and MemNet $+0.09$ to $+0.11$ at the $2\times$ norm; all bootstrap confidence intervals above zero). Second, $\boldsymbol{\theta}$ captures only a modest fraction of this reachable gain, and only for memorability. MemNet gains along $\boldsymbol{\theta}$ were $+0.007$ to $+0.031$ (confidence intervals above zero in all subjects), corresponding to 9\%--28\% of the iterative benchmark gain and 21\%--56\% of the single-step exact-gradient gain at the $1\times$ norm. In this benchmark, EmoNet steps along $\boldsymbol{\theta}$ moved scores in the unintended direction at both magnitudes. Per-image exact gradients were nearly orthogonal to $\boldsymbol{\theta}$ (mean cosine $0.00$--$0.04$). Third, a single exact-gradient step at the $2\times$ norm overshot for MemNet in all four subjects (mean change negative, confidence intervals below zero), whereas the same total norm applied iteratively did not. This benchmark demonstrates sensitivity to step size and update strategy under the tested norm constraints, providing a possible explanation for non-monotonic single-direction responses rather than establishing the mechanism of the human full-range effect.

Random unit directions produced negligible changes throughout. Together with Figure~\ref{fig:appendix_theta_validation}, these results clarify the interpretation of $\boldsymbol{\theta}$ used throughout the manuscript: a population-level attribute axis with a measured performance envelope, not a reliable proxy for every image's locally optimal direction. The analysis script and its outputs accompany the code described in Section~\ref{sec:data-code-availability}.

\section{Illustrative Reconstruction Sequences}
\label{app:reconstruction_examples}
\begin{figure}[t]
\centering
\includegraphics[width=0.99\textwidth]{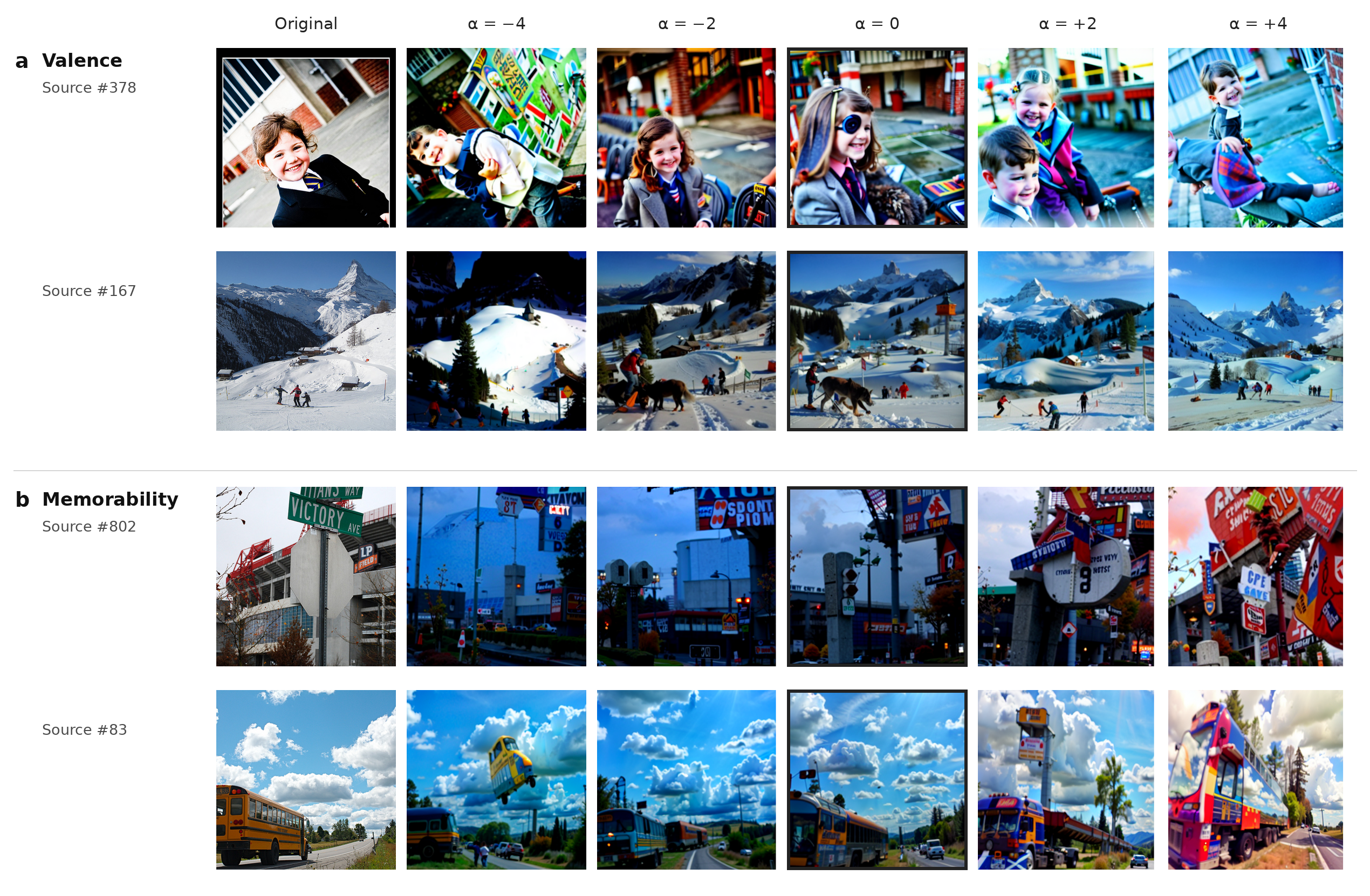}
\caption{\textbf{Illustrative valence and memorability perturbation sequences.}
\textbf{(a)} Two valence-targeted Versatile Diffusion reconstruction sequences from one fMRI subject.
\textbf{(b)} Two memorability-targeted sequences from the same reconstruction pipeline. The first column shows the original NSD image; the remaining columns show reconstructions at $\alpha \in \{-4, -2, 0, +2, +4\}$. The dark border identifies the unperturbed $\alpha=0$ reconstruction. Negative and positive values indicate the intended directions of decreasing and increasing automated attribute scores, respectively. The examples were selected for visibly interpretable trajectories and are not intended to represent the distribution of effects across all images. They illustrate both the intended modulation and the increasing generative drift observed at extreme perturbation strengths. Additional examples are available at \url{https://www.neuralsurrogate.com/}.}
\label{fig:appendix_exemplar_series}
\end{figure}

Figure~\ref{fig:appendix_exemplar_series} shows selected Versatile Diffusion reconstruction sequences across the five perturbation strengths. Two source images are shown for each target attribute. The examples were selected because they provide clearly visible and interpretable trajectories and should therefore be regarded as illustrative rather than random or statistically representative cases. They do not substitute for the group-level analyses reported in the main text.

The valence examples illustrate changes in color, illumination, social content, and scene composition across $\alpha$. The memorability examples show increasing visual distinctiveness, semantic density, or unusual object configurations at positive perturbation strengths. These sequences also make the limits of the method visible. At the extreme values of $\alpha$, particularly $\alpha=\pm4$, changes can extend beyond the targeted attribute and alter source-image identity, low-level appearance, or semantic content. The examples should consequently be interpreted together with the score and fidelity analyses in Figures~\ref{fig:Figure_4} and~\ref{fig:Figure_5}.

A larger gallery containing additional valence and memorability reconstruction sequences is available at \url{https://www.neuralsurrogate.com/}.

\FloatBarrier
\section{Surrogate-Predicted Neural Patterns Under Cognitive Perturbation}
\label{subsec:neural_patterns}
As a model-implied visualization, we examined how the latent-space perturbations analyzed in the main text are represented in the fMRI-derived surrogate. Predicted beta maps for valence and memorability at the extreme perturbation levels ($\alpha=\pm4$) were projected onto cortical surfaces using FreeSurfer~\citep{fischl2000measuring,fischl1999b} (Figure~\ref{fig:Figure_6}; computational details in Appendix~\ref{app:cortical_map_computation}). The maps show surrogate-predicted differences relative to the unperturbed condition ($\alpha=0$); they are not empirical activation maps obtained during perturbation and should not be interpreted as validated stimulation targets.
The resulting patterns were concentrated primarily within the visual and occipital regions represented in the surrogate. For valence, the $\alpha=-4$ condition showed larger apparent deviations from baseline across early visual and occipital cortex than the $\alpha=+4$ condition. For memorability, the negative perturbation produced a more spatially restricted pattern, whereas the $\alpha=+4$ condition was associated with broader predicted occipital changes. The descriptive similarity between the positive-valence and high-memorability maps may reflect overlap in the visual features emphasized by the two automated assessors rather than a shared neural mechanism.
These patterns must also be considered in relation to the fidelity results reported in Figure~\ref{fig:Figure_5}. Both $\alpha=-4$ and $\alpha=+4$ lie near the boundary at which reconstruction fidelity declined, with the strongest degradation generally occurring at the negative extreme. Consequently, the displayed differences may reflect a combination of the intended cognitive-attribute perturbation and representational changes associated with generative drift or image degradation. The maps therefore provide a qualitative visualization of how the surrogate translates latent perturbations into brain space, but they do not isolate the neural effect of the targeted attribute.
No inferential statistical tests were performed because the maps were generated by the model-based perturbation pipeline rather than estimated from independently measured neural responses under the corresponding conditions. Establishing whether these predicted patterns can be induced, whether they causally influence valence or memory, and whether they can inform stimulation design will require prospective experiments combining direct neural measurement with controlled perturbation or neurostimulation.
\begin{figure}[H]
\centering
\includegraphics[width=0.85\textwidth]{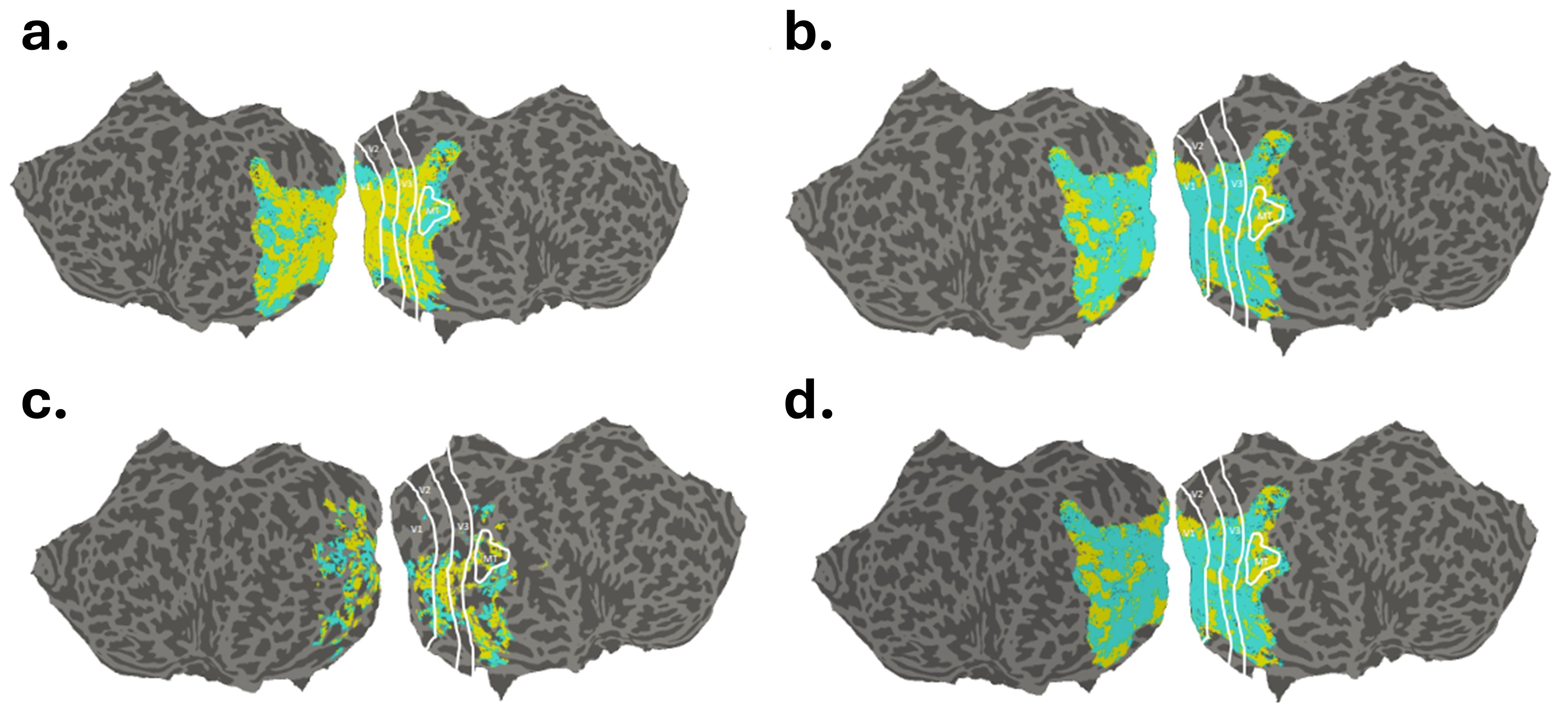}
\caption{\textbf{Surrogate-predicted neural patterns under extreme cognitive-attribute perturbations.}
Predicted cortical beta-pattern differences relative to the unperturbed condition ($\alpha=0$). Panels \textbf{a} and \textbf{b} show negative ($\alpha=-4$) and positive ($\alpha=+4$) valence perturbations, respectively; panels \textbf{c} and \textbf{d} show low- ($\alpha=-4$) and high-memorability ($\alpha=+4$) perturbations. Yellow indicates predicted increases and cyan indicates predicted decreases relative to $\alpha=0$. These maps are descriptive outputs of the fMRI-derived surrogate and do not represent independently measured activation changes or validated stimulation targets. No inferential statistical tests were performed.}
\label{fig:Figure_6}
\end{figure}

\FloatBarrier
\section{Behavioral Detrending and Sensitivity Analyses}
\label{app:behavioral_sensitivity}

\subsection{Rationale for the primary correction}
All participants completed the behavioral conditions in the same fixed block order: original NSD images, unperturbed reconstructions ($\alpha=0$), memorability blocks at $\alpha=-2,-4,+4,+2$, and valence blocks at $\alpha=-2,-4,+2,+4$. Consequently, condition and between-block time were not independently randomized. A difference between a perturbation block and the $\alpha=0$ block could therefore reflect the intended manipulation, differences in source-image composition, or a systematic change in rating behavior over the approximately 140-minute experiment.

\begin{figure}[htbp]
\centering
\includegraphics[width=0.99\textwidth]{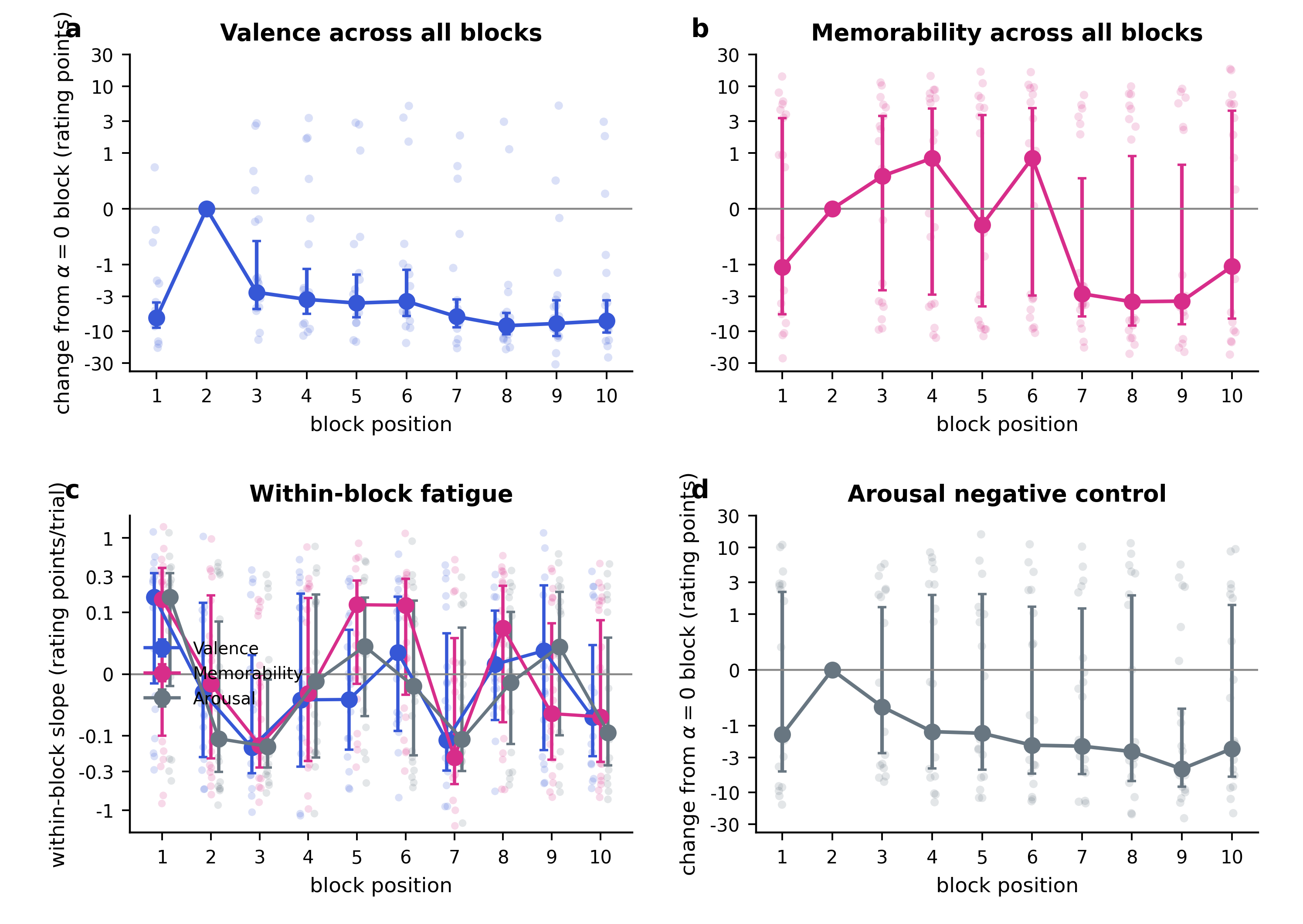}
\caption{\textbf{Behavioral time-course and negative-control diagnostics.}
\textbf{(a,b)} Changes in valence and perceived-memorability ratings across the ten fixed block positions, expressed relative to the unperturbed reconstruction block ($\alpha=0$, block 2). Small points show individual participants; lines connect only the group means, and error bars denote 95\% confidence intervals.
\textbf{(c)} Participant-specific within-block rating slopes for valence, perceived memorability, and arousal. Slopes were estimated across the 40 trial positions within each block.
\textbf{(d)} Arousal-rating changes across block position, used as a negative-control outcome because arousal was not directly targeted. The between-block curves are descriptive rather than causal estimates because block position, image composition, and perturbation condition were not independently randomized.}
\label{fig:appendix_behavioral_diagnostics}
\end{figure}

The uncorrected ratings showed substantial between-block variation, particularly for valence (Figure~\ref{fig:appendix_behavioral_diagnostics}a). By contrast, participant-specific slopes within individual blocks were generally small and centered near zero for valence, memorability, and arousal (Figure~\ref{fig:appendix_behavioral_diagnostics}c). This pattern suggests that the principal temporal nuisance occurred across blocks rather than as a continuous decline within each 40-trial block. Arousal, which was not explicitly targeted by the perturbations, showed no systematic full-range effect. Nevertheless, the between-block curves remain descriptive because the images and target conditions also differed across blocks. They demonstrate the need for sensitivity analysis but cannot, by themselves, identify a unique fatigue function.

The primary analysis therefore used a leave-one-participant-out, assessor-neutral linear correction. For target $k$, let $y_{pij}$ denote the rating of participant $p$ for source image $i$ in block $j$. The nuisance model was
\begin{equation}
y_{pij} =
\gamma_p
+
\beta_1 b_i
+
\beta_2 b_i^2
+
\beta_3 w_{ij}
+
\beta_4 t_j
+
\varepsilon_{pij},
\label{eq:behavioral_nuisance_linear}
\end{equation}
where $\gamma_p$ denotes fixed effects for the training participants, $b_i$ is the standardized assessor score of the source image in the unperturbed $\alpha=0$ condition, $w_{ij}$ is centered within-block trial position divided by 10, and $t_j$ is block position relative to the $\alpha=0$ block. The quadratic term in $b_i$ allows the association between baseline image content and human ratings to be nonlinear.

To reduce contamination of the temporal estimate by the intended perturbation, Equation~\ref{eq:behavioral_nuisance_linear} was fitted only to the complete $\alpha=0$ condition and to perturbed images for which the absolute Versatile Diffusion assessor change was no larger than 0.50 baseline SD. These images were treated as assessor-neutral for estimation of nuisance trends. For each held-out participant, all nuisance coefficients were estimated using the remaining 17 participants; the held-out participant therefore contributed no data to their own correction.

The corrected rating was
\begin{equation}
y_{pij}^{\mathrm{corr}} =
y_{pij} -
\left(
\widehat{\beta}_1 b_i
+
\widehat{\beta}_2 b_i^2
+
\widehat{\beta}_3 w_{ij}
+
\widehat{\beta}_4 t_j
\right).
\label{eq:behavioral_correction}
\end{equation}
Participant fixed effects were used during nuisance-model fitting but were not subtracted from the held-out observations. For each participant and perturbation level, the corrected condition mean was compared with that participant's corrected $\alpha=0$ mean and divided by the participant-specific SD of the $\alpha=0$ ratings. The resulting estimand is therefore a within-participant change expressed in baseline-SD units.

This correction was chosen because it is the minimum-complexity model that simultaneously addresses source-image composition, within-block position, and the dominant between-block trend. The use of assessor-neutral observations and leave-one-participant-out estimation limits, but cannot eliminate, dependence on the fixed experimental order. The corrected estimates must therefore remain conditional on the stated time model.

\subsection{Sensitivity to the correction specification}
\begin{figure}[htbp]
\centering
\includegraphics[width=0.99\textwidth]{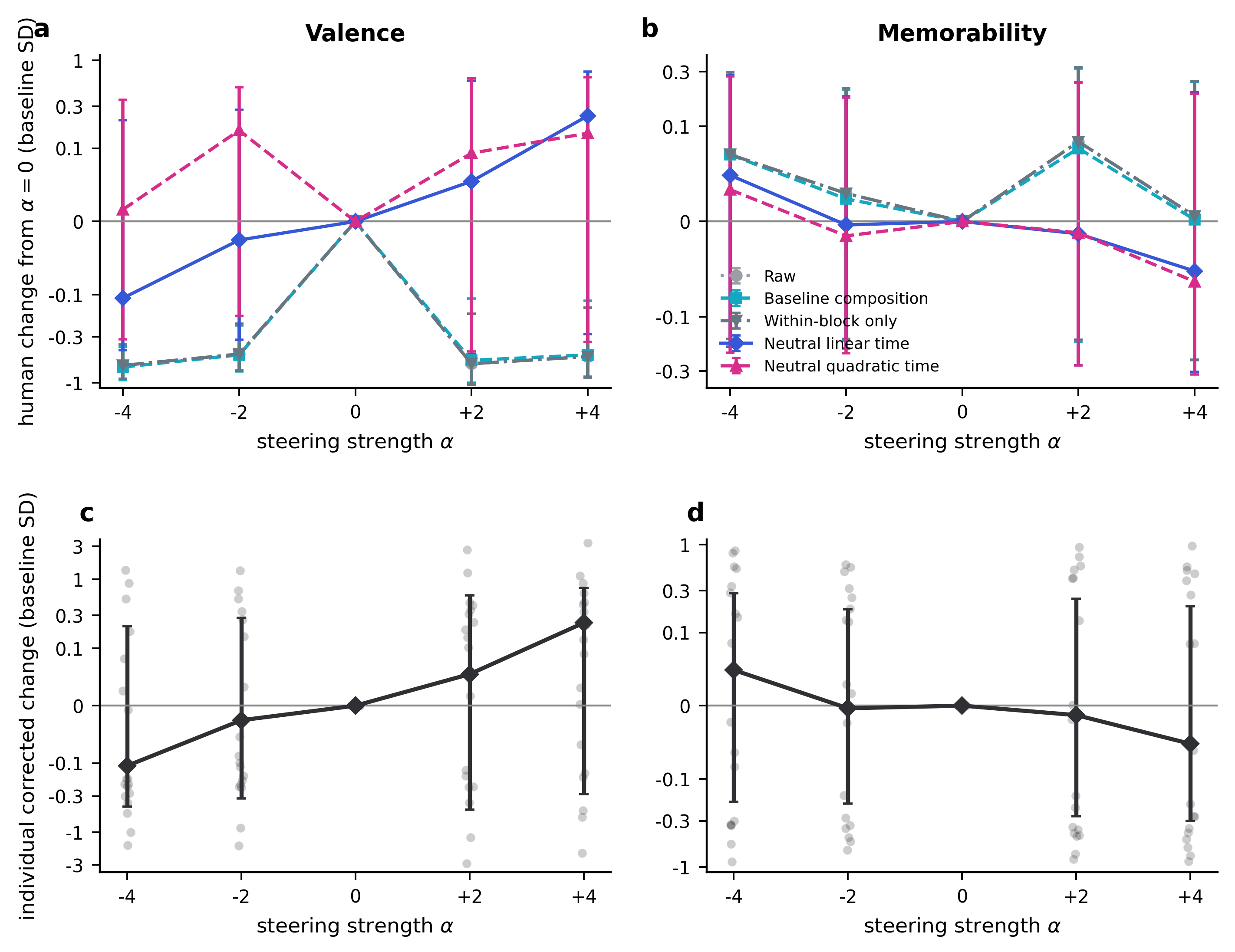}
\caption{\textbf{Sensitivity of the human effects to alternative detrending specifications.}
\textbf{(a,b)} Mean valence and perceived-memorability changes under the raw, baseline-composition, within-block-only, assessor-neutral linear-time, and assessor-neutral quadratic-time specifications. Changes are expressed relative to $\alpha=0$ in participant-specific baseline-SD units; error bars denote 95\% confidence intervals.
\textbf{(c,d)} Participant-level observations after the primary leave-one-participant-out assessor-neutral linear correction. Small points are unconnected participant estimates; lines join only group means. The comparison shows that the valence inference depends principally on correction of the between-block trend, whereas perceived memorability remains close to zero under every specification.}
\label{fig:appendix_correction_sensitivity}
\end{figure}

Figure~\ref{fig:appendix_correction_sensitivity} compares five correction specifications, with full-range slope summaries in Table~\ref{tab:behavioral_slope_sensitivity}: no correction; correction for baseline source-image composition; correction for within-block position alone; assessor-neutral linear time; and assessor-neutral quadratic time. The raw, composition-only, and within-block-only valence estimates remained substantially negative across most conditions because these specifications did not account for the pronounced between-block shift. Adjusting for within-block position alone had almost no effect, consistent with the near-zero within-block slopes in Figure~\ref{fig:appendix_behavioral_diagnostics}c.

The assessor-neutral linear model removed most of the common downward displacement and produced the positive full-range valence contrast reported in the main text. The mean participant slope from $\alpha=-4$ to $+4$ was 0.038 baseline SD per unit of $\alpha$ (95\% CI $[0.003, 0.074]$); 16 of 18 participant slopes were positive. The Holm-adjusted exact binomial test was $p=0.0039$, whereas the Holm-adjusted sign-flip test of the mean slope was borderline ($p=0.0527$). Only one participant showed a strictly increasing sequence across every adjacent $\alpha$-level. The main result therefore supports an overall positive directional valence contrast, not a uniformly monotonic participant-level dose--response.

A quadratic between-block model was retained as the principal time-model sensitivity analysis:
\begin{equation}
y_{pij} =
\gamma_p
+
\beta_1 b_i
+
\beta_2 b_i^2
+
\beta_3 w_{ij}
+
\beta_4 t_j
+
\beta_5 t_j^2
+
\varepsilon_{pij}.
\label{eq:behavioral_nuisance_quadratic}
\end{equation}
At the primary 0.50-SD neutrality threshold, held-out-image prediction modestly favored the quadratic model for valence ($\Delta\mathrm{RMSE}_{\mathrm{Q}-\mathrm{L}} = -0.050$, 95\% CI $[-0.099, -0.005]$). However, held-out-image folds retain the same block positions in their training and test sets and therefore primarily evaluate interpolation within the observed order. The more relevant leave-one-block-out analysis favored the linear model: quadratic-minus-linear RMSE was $+5.19$ for valence and $+0.03$ for memorability, with positive values indicating worse quadratic prediction. BIC also favored the linear model at the primary threshold ($\Delta\mathrm{BIC}_{\mathrm{Q}-\mathrm{L}} = +1.02$ for valence and $+4.87$ for memorability).

The same pattern was observed across assessor-neutrality thresholds of 0.25, 0.50, and 0.75 SD. Leave-one-block-out RMSE differences were consistently positive for valence ($+5.19$ to $+7.42$) and memorability ($+0.03$ to $+0.43$), and BIC differences consistently favored the linear model (valence: $+1.02$ to $+2.50$; memorability: $+2.57$ to $+5.08$). Bootstrap intervals for the quadratic coefficient included zero under every target-by-threshold specification. These diagnostics support the linear model as the more parsimonious and transportable primary correction.

Nevertheless, the valence effect was attenuated under quadratic correction: the full-range slope was 0.010 SD per unit of $\alpha$ (95\% CI $[-0.020, 0.040]$), with 11 of 18 slopes positive. Thus, the positive valence result is not invariant to the assumed shape of the between-block trajectory. This sensitivity is why the main text describes the human result as preliminary and conditional on the linear fatigue model. In contrast, the absence of a perceived-memorability effect was stable across all correction specifications.

\begin{table}[htbp]
\centering
\scriptsize
\caption{\textbf{Full-range participant-slope sensitivity across behavioral analyses.}
Slopes were estimated across $\alpha = -4, -2, 0, +2, +4$ and are expressed in participant-specific baseline SD per unit of $\alpha$. Confidence intervals summarize the participant-level slopes. Exact-source slopes were calculated from the corresponding participant-specific matched-image condition means.}
\label{tab:behavioral_slope_sensitivity}
\begin{tabular}{lcccc}
\toprule
& \multicolumn{2}{c}{Valence} & \multicolumn{2}{c}{Perceived memorability} \\
\cmidrule(lr){2-3}\cmidrule(lr){4-5}
Analysis
& Mean slope [95\% CI] & Positive
& Mean slope [95\% CI] & Positive \\
\midrule
Raw
& $\phantom{-}0.006$ [$-0.021$, $\phantom{-}0.034$] & 11/18
& $-0.004$ [$-0.021$, $\phantom{-}0.013$] & 7/18 \\
Baseline composition
& $\phantom{-}0.014$ [$-0.015$, $\phantom{-}0.044$] & 12/18
& $-0.004$ [$-0.021$, $\phantom{-}0.013$] & 7/18 \\
Within-block only
& $\phantom{-}0.006$ [$-0.021$, $\phantom{-}0.034$] & 11/18
& $-0.004$ [$-0.021$, $\phantom{-}0.013$] & 7/18 \\
Assessor-neutral linear (primary)
& $\phantom{-}0.038$ [$\phantom{-}0.003$, $\phantom{-}0.074$] & 16/18
& $-0.011$ [$-0.028$, $\phantom{-}0.007$] & 5/18 \\
Assessor-neutral quadratic
& $\phantom{-}0.010$ [$-0.020$, $\phantom{-}0.040$] & 11/18
& $-0.009$ [$-0.027$, $\phantom{-}0.008$] & 6/18 \\
Exact-source matched
& $\phantom{-}0.026$ [$-0.008$, $\phantom{-}0.061$] & 13/18
& $-0.007$ [$-0.026$, $\phantom{-}0.011$] & 5/18 \\
Exact-source + assessor-neutral
& $\phantom{-}0.037$ [$\phantom{-}0.004$, $\phantom{-}0.071$] & 12/18
& $-0.002$ [$-0.033$, $\phantom{-}0.030$] & 7/18 \\
\bottomrule
\end{tabular}
\end{table}

\subsection{Exact-source matched sensitivity analysis}
Differences in source-image composition were examined more directly by pairing every nonzero-$\alpha$ image with the identical source image presented in the $\alpha=0$ block. This analysis eliminates between-condition variation in source identity. The leave-one-participant-out linear coefficients were still used to adjust for the time and within-block differences between the paired presentations because exact source matching alone cannot resolve the fixed-order confound.

\begin{figure}[htbp]
\centering
\includegraphics[width=0.99\textwidth]{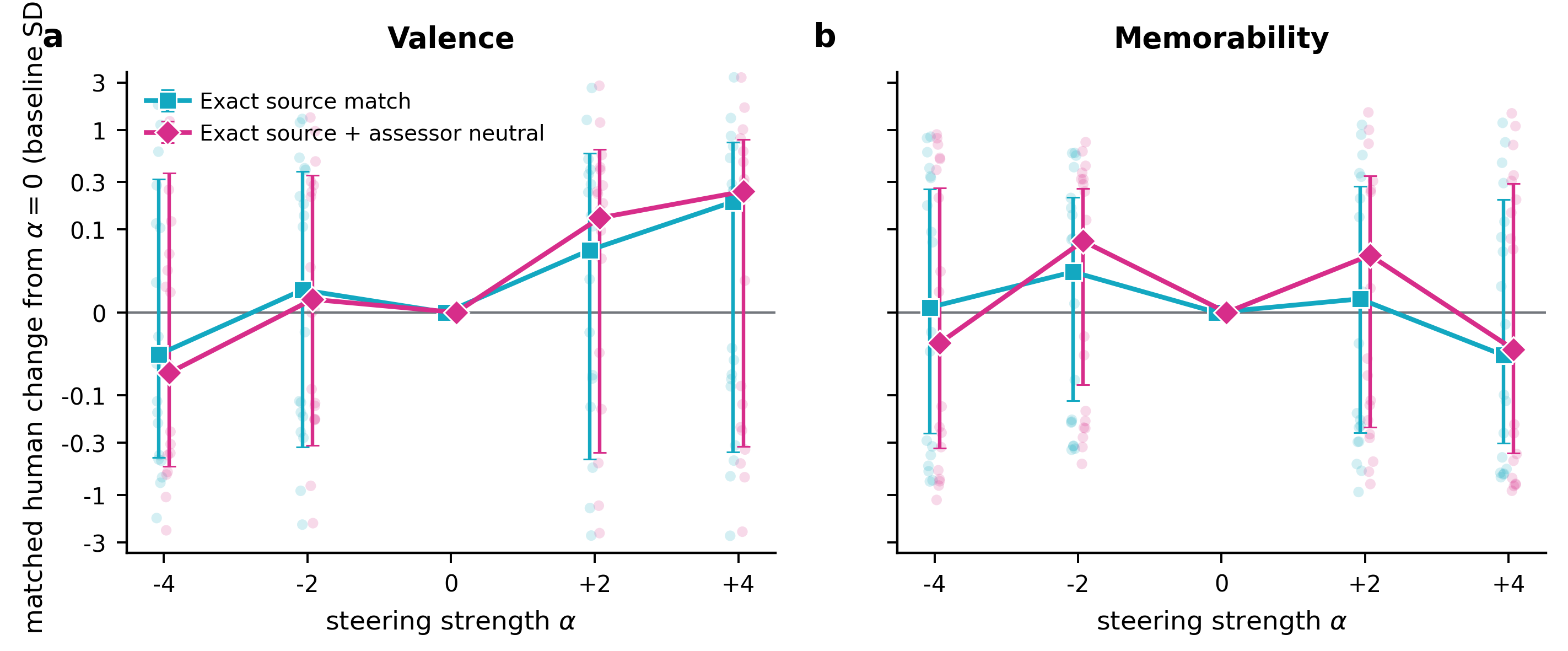}
\caption{\textbf{Exact-source matched behavioral sensitivity analysis.}
\textbf{(a)} Valence changes when each nonzero-$\alpha$ image was paired with the same source image in the $\alpha=0$ block.
\textbf{(b)} Corresponding analysis for perceived memorability. Cyan squares show all available exact-source pairs; magenta diamonds additionally restrict the analysis to assessor-neutral images with an absolute Versatile Diffusion assessor change no larger than 0.50 baseline SD. Small points show individual participants; lines connect only group means, and error bars denote 95\% confidence intervals. Values are displayed on a signed symmetric-logarithmic scale. Exact matching controls source-image identity but does not remove the fixed-order confound; the leave-one-participant-out linear time correction was therefore retained.}
\label{fig:appendix_exact_source}
\end{figure}

Two matched subsets were examined (Figure~\ref{fig:appendix_exact_source}). The first retained every available exact-source pair. The second additionally required the Versatile Diffusion assessor change to remain within 0.50 baseline SD of zero, providing an assessor-neutral matched subset. Exact matching reduced the number of images contributing to each participant-level estimate and therefore produced wider confidence intervals.

The exact-source results supported the qualitative distinction between the two behavioral targets. For valence, the mean full-range slope was positive when all available matched images were included (0.026 SD per unit of $\alpha$, 95\% CI [$-0.008, 0.061$]) and closely matched the primary estimate when both exact-source and assessor-neutral restrictions were applied (0.037, 95\% CI [$0.004, 0.071$]). The corresponding primary estimate was 0.038. This agreement indicates that the positive valence direction under the linear model was not produced solely by differences in source-image composition.

The exact-source analyses likewise provided no clear evidence of a perceived-memorability effect. The full-range slopes were $-0.007$ (95\% CI [$-0.026, 0.011$]) for all matched images and $-0.002$ (95\% CI [$-0.033, 0.030$]) for the assessor-neutral matched subset. The lack of clear evidence for a human perceived-memorability effect was therefore consistent across the tested correction methods and source-matching analyses.

Taken together, these analyses delineate the assumptions underlying the reported behavioral interpretation. Composition-only and within-block adjustments did not address the dominant fixed-order shift, motivating an explicit model-based correction for between-block time. The assessor-neutral linear model was selected as the primary analysis because it was cross-fitted, parsimonious, and favored by BIC and held-out-block prediction. Exact-source matching retained a positive valence estimate, with a nearly identical magnitude in the assessor-neutral subset, whereas perceived memorability remained consistently inconclusive. At the same time, the attenuation under quadratic detrending demonstrates that the valence result remains model-dependent. The behavioral evidence should therefore be interpreted as preliminary support for an overall positive valence direction under the stated primary correction, rather than as a model-invariant or uniformly monotonic dose--response.

\FloatBarrier
\section{Image-Quality Sensitivity Analysis}
\label{app:qc_sensitivity}

The primary analyses excluded no image. To assess how the conclusions depend on image-quality filtering, we prespecified and version-controlled, before computation, a uniform two-metric rule: an image is flagged only if both PixCorr and SSIM \citep{wang2004image} relative to its own unperturbed reconstruction fall below $\tau=0.5$, applied identically to every reconstruction stage, target, fMRI subject, and perturbation level. The rule was computed for all 62{,}848 perturbed images (four subjects, two stages, two targets, four nonzero perturbation levels, and 982 held-out images each), and the full-range analyses and restricted-range sensitivity analyses were repeated where estimable under the frozen rule and a threshold grid $\tau\in\{0.3, 0.4, 0.5, 0.6, 0.7\}$. The criteria use only stimulus fidelity; assessor scores and behavioral outcomes play no role.

Pooled across subjects, the frozen rule flagged 0.1\%--0.5\% of VDVAE images at the intermediate perturbation levels ($\alpha=\pm2$) and 4\%--11\% at $\alpha=\pm4$. For Versatile Diffusion it flagged 56\%--75\% of images at the intermediate perturbation levels ($\alpha=\pm2$) and 77\%--95\% at the extremes. Low-level similarity metrics alone cannot establish perceptual validity for diffusion-refined images, whose texture can be redrawn during generation. Fidelity metrics therefore do not define exclusions for the main analysis; filtering is used only for sensitivity checks.

In the intermediate-level route-comparison subset, the direct-route automated results were unchanged under the frozen rule: no VDVAE image was excluded from that subset, and the paired contrasts and monotonicity classifications were identical. A threshold-dependent automated result was the net score advantage of the neural route over direct steering (Appendix~\ref{role_of_R}). This advantage was sensitive to retaining the images that routing altered most: at $\tau=0.4$, fewer than a quarter of the neural-routed valence images and about half of the memorability images remained, and the differences became inconsistent in sign; at $\tau\geq0.5$, too few neural-routed images remained for estimation.

For the human data, participants rated a single Versatile Diffusion reconstruction for each source-image--condition combination, but the behavioral files did not record which fMRI subject supplied that reconstruction. Flags were therefore computed from the subject-averaged Versatile Diffusion fidelity metrics, and unperturbed trials were never flagged. The participant-slope analysis of the main text was repeated after removing flagged trials. The subject-averaged metrics indicated the greatest degradation at $\alpha=-4$; the rule removed 80\% of the rated valence images and 60\% of the rated memorability images at $\alpha=-4$ already at $\tau=0.3$, and every rated valence image at $\alpha=-4$ at $\tau\geq0.5$, where the full-range analysis is consequently infeasible. The restricted-range sensitivity slopes across $\alpha=-2,0,+2$ remained statistically inconclusive at every feasible threshold (valence: 0.019 and 0.042 SD per unit of $\alpha$ at $\tau=0.3$ and $0.4$; memorability: 0.009 and 0.037; all confidence intervals included zero). For the full-range valence analysis, the confidence intervals included zero after filtering: at $\tau=0.3$ the slope was 0.035 (95\% CI [$-0.004$, 0.075]), positive in 12 of 18 participants (Holm-adjusted binomial $p=0.71$), and at $\tau=0.4$ it was $-0.004$ (95\% CI [$-0.052$, 0.044]). The human perceived-memorability slopes remained statistically inconclusive at every feasible threshold. These results show that the full-range valence estimate is sensitive to the retained stimulus set. At $\tau=0.3$, the point estimate remained close to the unfiltered value of 0.038, but its interval widened to include zero; at $\tau=0.4$, the estimate was attenuated. At $\tau\geq0.5$, removal of all rated valence source images at $\alpha=-4$ made the original five-level estimand unavailable. The flags are based on source-index-matched, subject-averaged fidelity rather than the exact displayed reconstruction. Consequently, these checks qualify robustness but do not establish that degradation caused the original effect. The main analysis remains the unfiltered full-range comparison, conditional on the time-correction model documented in Appendix~\ref{app:behavioral_sensitivity}.

The manifest and the route and human-slope sensitivity tables accompany their generating scripts in the code described in Section~\ref{sec:data-code-availability}.

\end{appendices}


\end{document}